\documentclass{aa}

\usepackage{graphicx}
\usepackage{txfonts}
\usepackage{color}
\usepackage{natbib}
\newcommand{\SC}[1]{\textcolor{black}{#1}} 
\newcommand{\BT}[1]{\textcolor{black}{#1}}
\newcommand{\kms}{km~s$^{-1}$}

\bibpunct{(}{)}{;}{a}{}{,} 

\begin{document}

   \title{Interaction between a pulsating jet and a surrounding disk wind}

  \subtitle{A hydrodynamical perspective}

   \author{B. Tabone 
        \inst{1} 
         A. Raga \inst{2},
        S. Cabrit \inst{1, 3}, G. Pineau des For\^ ets \inst{1, 4}
            }

 \institute{LERMA, Observatoire de Paris, PSL Research University, CNRS, Sorbonne Universit\'e, UPMC Univ. Paris 06, 75014 Paris, France 
                       \and
                       Instituto de Ciencias Nucleares, Universidad Nacional Autónoma de México, Ap. 70-543, 04510 Cd. Mx., Mexico 
                       \and
             Univ. Grenoble Alpes, CNRS, IPAG, 38000 Grenoble, France
                       \and
             Institut d’Astrophysique Spatiale, CNRS UMR 8617, Université Paris-Sud, 91405 Orsay, France
       }

\abstract
{The molecular richness of fast protostellar jets within 20-100 au of their source, despite strong ultraviolet irradiation, remains a challenge for the models investigated so far.}
{We aim \BT{to investigate} the effect of interaction between a time-variable jet and a surrounding steady disk wind, 
to assess the possibility of jet chemical enrichement by the wind, and the characteristic signatures of such a configuration.}
{We \BT{have constructed} an analytic model 
of a jet bow shock 
driven into a surrounding slower disk wind in {the thin shell approximation}.
The refilling of the post bow shock cavity from below by the disk wind is also studied. An extension of the model to the case
of two or more successive internal working surfaces (IWS) is made.
We then \BT{compared} this analytic model with numerical simulations with and without a surrounding disk wind.} {We find that at early times (of order the variability period), jet bow shocks travel in refilled pristine disk-wind material, before interacting with the cocoon of older bow shocks. This opens the possibility of bow shock chemical enrichment (if the disk wind is molecular and dusty) and of probing the unperturbed disk wind structure near the jet base. 
Several distinctive signatures of the presence of a surrounding disk wind are identified, in the bow shock morphology and kinematics. Numerical simulations validate our analytical approach and further show that at large scale, the passage of many jet IWS inside a disk wind produces a stationary V-shaped cavity, closing down onto the axis at a finite distance from the source.}{}


   \maketitle

\section{Introduction}

Protostellar jets appear intimately linked to the process of mass accretion onto the growing star; their strikingly similar properties across protostellar age,  mass, and accretion rate all point to universal ejection and collimation mechanisms
\citep{2002EAS.....3..147C,2007A&A...468L..29C, 2013A&A...551A...5E}. Yet, jets from the youngest protostars --- so-called Class 0 --- are much brighter in molecules \citep[\BT{e.g.,}][]{2000A&A...359..967T} than jets from more evolved protostars and pre-main sequence stars which are mainly atomic; Molecules have been traced as close as 20-100 au from the source \citep[\BT{e.g.,}][]{2017NatAs...1E.152L,2014ApJ...794..169H}. 
The origin of this selective molecular richness remains an important issue for models of the jet origin.
Three broad scenarios have been considered, with no fully validated answer so far. 

In models of ejection from the stellar magnetosphere or the inner disk edge \citep[\BT{e.g.,}][]{1994ApJ...429..781S,2002ApJ...578..420R,2005ApJ...632L.135M,2013A&A...550A..99Z}, the jet
would be expected to be dust-free (the grain sublimation radius around a typical solar-mass protostar is $R_{\rm sub}\sim 0.3$~au,  see \BT{for example} \citet{2016A&A...585A..74Y}).
The lack of dust screening then makes the wind extremely sensitive to photodissociation by the accretion shock.
Chemical models of dust-free winds by \citet{1991ApJ...373..254G} found that CO, \BT{SiO,} and H$_2$O could no longer form at the wind base in the presence of a typical expected level of FUV excess\footnote{the flat UV flux in their UV1-UV2 models is $\sim$ 30-500 that in BP Tau \citep{2003ApJ...591L.159B}, for a wind-mass flux corresponding to a 1000 times larger accretion rate ($\sim 3 \times 10^{-5} M_\odot$ yr$^{-1}$)}. \citet{2005RMxAA..41..137R} showed that H$_2$ could form further out behind internal shocks. However, the key ions involved are also easily destroyed by FUV photons. 
Hence, molecule formation in a dust-free jet within 20-100 au of protostars remains an open issue.

A second proposed explanation 
is that the molecular component of jets may be tracing dusty MHD disk winds launched beyond $R_{\rm sub}$, where dust can shield molecules against the FUV field and allow faster H$_2$ reformation. Detailed models are successful at reproducing the higher molecule richness of Class 0 jets \citep{2012A&A...538A...2P}
the broad water line components revealed by \textit{Herschel}/HIFI  \citep{2016A&A...585A..74Y}, and the rotation signatures \BT{recently resolved by ALMA in the HH212 jet and in the slow wider angle wind surrounding it} \citep{2017NatAs...1E.152L, 2017A&A...607L...6T}. However, the same disk wind models predict that the fastest, SiO-rich streamlines in  HH212 (flowing at $\sim 100$ \kms)  would be launched
from 0.05-0.2 au, within the dust sublimation radius \citep{2017A&A...607L...6T}. Hence, this scenario still partly faces the unsolved question of molecule survival in a dust-free wind.

A third scenario is that molecules could be somehow "entrained" from the surroundings into the jet, assumed initially atomic. In a time-dependent jet, travelling internal shocks will squeeze out high-pressure jet material, which then sweeps up the surrounding gas into a curved bowshock. If the surrounding material is molecular, a partly molecular bowshock will result, with a more tenuous "wake" of shocked molecular gas trailing behind it \citep{1993A&A...278..267R,2005RMxAA..41..137R}. As the next "internal working surface" (IWS) propagates into this wake, it may again produce a molecular jet bowshock. However,  after the passage of many such IWS, the wake will be so shock-processed and tenuous that not enough molecules may be left to produce molecular bowshocks close to the jet axis.

In the present paper, we revisit this last scenario in a new light by investigating whether a slower molecular "disk wind" surrounding the jet could help refill the wake and re-inject
fresh (unprocessed) molecules into the jet path. This new outlook is prompted by the discovery of a potential molecular disk wind wrapped around the dense axial jet in HH212 \citep{2017A&A...607L...6T}. We explore this possibility by studying analytically and numerically the propagation of bow shocks driven by a time-variable, inner jet into a surrounding slower disk wind. This scenario may be seen as an extension of the recent modeling work of \citet{2016MNRAS.455.2042W} who studied the turbulent mixing layer between a jet and disk wind, with the novel addition of internal working surfaces in the jet to produce a stronger coupling between the two outflow components. 

Besides our main goal of exploring the impact of a DW on the chemical richness of Class 0 jets, our study has two other important motivations. \BT{First, we aim to identify specific signatures in the morphology and kinematics of jet bowshocks that could reveal the presence of a surrounding DW. Secondly, we aim to
identify in which regions of space the pristine DW material would remain unperturbed, for comparison to theoretical DW models.}

In the present exploratory study, \BT{we have limited} ourselves to purely hydrodynamical and cylindrical flows, which allow us to develop an analytical model that greatly helps to capture the main effects of the two-flow interaction, and to understand the numerical results. Also, this is expected to be an optimal case for interaction between the two flows, as magnetic tension would tend to oppose mixing. 

The paper is organized as follows.
In \BT{Section 2}, we build an analytical model (in the thin shell approximation) for the propagation of a bow shock driven by an IWS into
a surrounding disk wind. The model is extended to the case of two or
more successive IWS in \BT{Section 3}. In \BT{Section 4} we compare the analytic model with axisymmetric
simulations of a variable jet+surrounding disk wind configuration, and compare the results with a
"reference simulation" in which the same variable jet propagates into a stationary environment.
Finally, the results are discussed in \BT{Section 5}.

\section{Analytical approach}
\subsection{Basic equations}

\begin{figure}[!t]
\centering
\includegraphics[width=8cm]{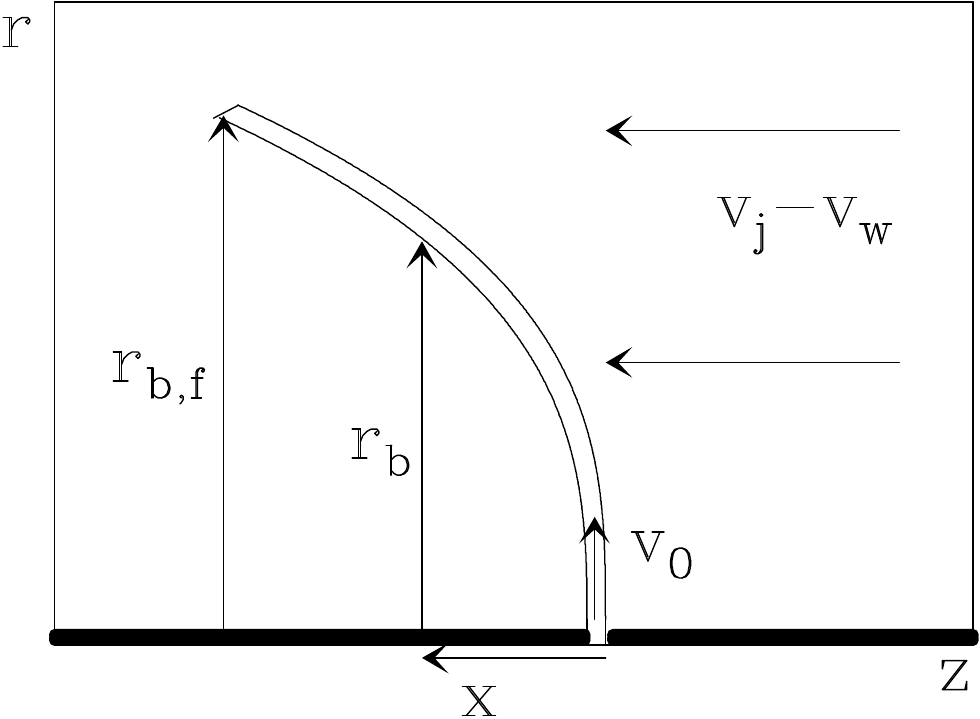}
\caption{Schematic diagram of the flow around an internal working surface (IWS)
in the frame of reference co-moving with the IWS
at velocity v$_z$ = v$_j$ (a similar
configuration would apply for the leading working surface of a jet).
The thick, horizontal line at the bottom of the graph is the jet
(with a gap showing the position of the IWS).
The working surface ejects jet material sideways at an initial velocity
v$_0$ into the slower disk wind, which in this frame of reference moves
towards the outflow source at velocity v$_j-$v$_w$. The distance $x$
is measured towards the outflow source.
The shape of the thin shell bow shock 
is given by $r_b(x)$ (see Equ.~\ref{rx}), and terminates 
at the cylindrical radius $r_{b,f}(t)$ with $t$ the time elapsed since formation of the IWS
(see Equ.~\ref{rf}).
}
\label{fig:shockframe}
\end{figure}

We \BT{considered} the "disk wind+jet" configuration shown in \BT{Fig.}~\ref{fig:shockframe}.
where a  
cylindrical jet of radius $r_j$ and time-variable velocity v$_j$ directed along the $z$-axis
is immersed in a plane-parallel "disk wind'' 
with uniform density $\rho_w$ and time-independent velocity v$_w$ parallel to v$_j$.

\begin{center}
\begin{figure*}[!t]
\centering
\includegraphics[width=12cm]{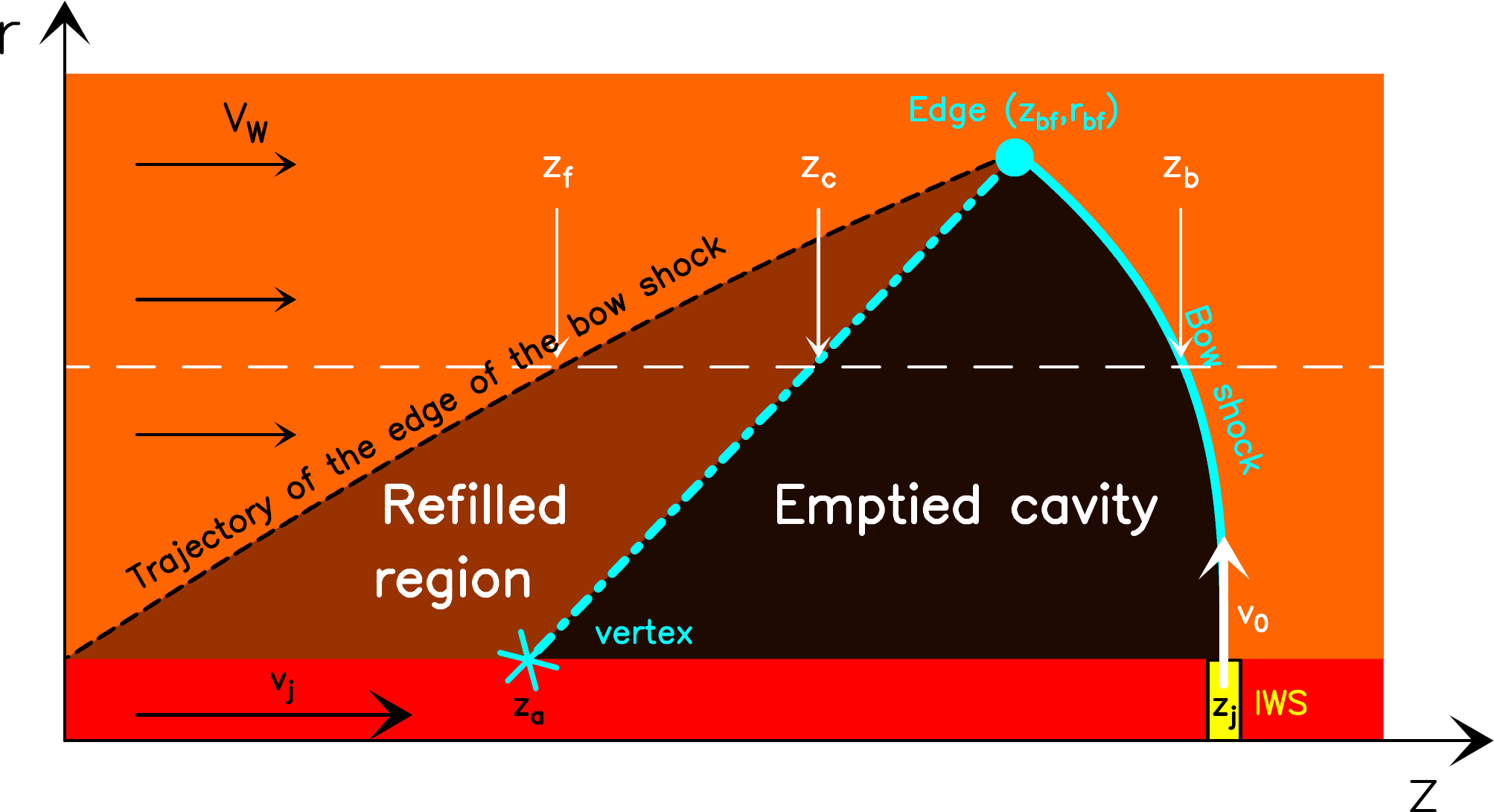} 
\caption{Schematic diagram showing the flow around a working surface of a jet
(in this case the
leading working surface, but the diagram also applies for an internal
working surface). 
The jet is the horizontal, red rectangle at the bottom of the graph, with the source located at $z$=0. 
The working surface in yellow is located at a distance $z_j$ from the source and travels
at a velocity v$_j$.  It ejects material away from the axis at an initial velocity v$_0$. The jet is
surrounded by a ``disk wind'', which travels along the outflow
axis at a velocity v$_w$.  The shape of the thin-shell bow shock  (thick cyan line) is given by
$z_b$ as a function of $r$ and ends at the edge of the bow wing (cyan point) (z$_{bf}$,r$_{bf}$). 
The bow shock leaves behind a ``cavity'' (black region), which is partially refilled by the disk wind (brown region). 
The boundaries of the initial swept up cavity (in black dashed line) 
and of the refilled region (cyan dash-dotted line) are given by $z_f$ an $z_c$ (respectively) as
a function of cylindrical radius $r$ (see \BT{Eqs.}~\ref{zfrf2} and \ref{zc}).
}
\label{fig:restframe}
\end{figure*}
\end{center}

The jet velocity variation is such that 
an internal working surface is produced within the jet beam. 
In the following derivations, we assume
that the working surface is formed at $t=0$ at the position of the 
source (\BT{i.e.,} $z=0$), 
and that it then travels at a constant
velocity v$_j$ (for $t>0$). Such a working surface could be produced,
\BT{for example}, by an outflow velocity with a constant value v$_1<$v$_j$ for $t<0$,
jumping to a constant value v$_2>$v$_j$ for $t\geq 0$.
Note that if the shock is produced at distance $z_s>0$ at a time $t_s >0$, 
the equations below remain valid with the transformation $z\rightarrow z-z_s$ and $t\rightarrow t-t_s$. 

In a frame of reference moving with the internal working surface  \SC{(see \BT{Fig.}~\ref{fig:shockframe})}
the over-pressured shocked jet material which is ejected sideways from the
working surface interacts with the slower moving, surrounding disk wind.
In the strong radiative cooling limit, this sideways ejection leads to
the formation of a thin-shell bow shock, which sweeps up material of the
surrounding disk wind, 
flowing towards the source at a relative velocity (v$_j$-v$_w$).

Assuming full mixing between jet and disk-wind material, we can write the mass, $r$-
and $x$-momentum conservation equations 
at any point of radius $r_b$ along this thin-shell ($r$,  $x$ and $r_b$ being defined in \BT{Fig.}~\ref{fig:shockframe})
as:
\begin{equation}
{\dot m}={\dot m}_0+\int_{r_j}^{r_b} 2\pi r'\rho_w(\textnormal{v}_j-\textnormal{v}_w)dr'\,,
\label{mass}
\end{equation}
\begin{equation}
{\dot \Pi}_r={\dot m}_0 \textnormal{v}_0={\dot m}\textnormal{v}_r\,,
\label{rmom}
\end{equation}
\begin{equation}
{\dot \Pi}_x=\int_{r_j}^{r_b} 2\pi r'\rho_w(\textnormal{v}_j-\textnormal{v}_w)^2dr'={\dot m} \textnormal{v}_x\,,
\label{xmom}
\end{equation}
where 
${\dot m}$ is the mass rate, ${\dot \Pi}_r$ the $r$-momentum rate and
${\dot \Pi}_x$ the $x$-momentum rate of the \SC{mixed jet+disk-wind} material flowing along the thin-shell
bow shock, and ${\dot m}_0$ and $v_0$ are the mass rate and velocity (respectively)
\SC{at which jet material is initially} ejected sideways by the working surface. These equations have
straightforward interpretations. As an illustration, we point out that \BT{Eq.}~(\ref{rmom})
states that the radial momentum of the material flowing along the thin shell remains constant over time (which is due to the fact that the disk wind material adds no $r$-momentum), so that
its radial velocity v$_r$ decreases as \SC{${\dot m}$ increases,} the $r$-momentum being shared with \SC{a larger amount of}
zero $r$-momentum material from the disk wind. 

The mass rate ${\dot m}_0$ and velocity v$_0$ of the sideways
ejected material are determined \SC{only} by the properties of the working surface. For
a highly radiative working surface, we would expect the \SC{post-shock jet}
material to cool to $\sim 10^4$~K before exiting sonically into the disk
wind. Therefore, we would expect v$_0\sim 10$~km~s$^{-1}$. 
The mass rate  ${\dot m}_0$ 
will have values of the order of the (time-dependent) mass loss rate ${\dot M}_j$
\SC{in} the jet beam. 

\SC{We note that although our basic equations are similar to those of  \citet{2001ApJ...557..443O} our approaches and derived equations will differ.
They considered only the case of the leading jet bowshock propagating in a medium at rest (v$_w = 0$), so that
the injected mass and momentum rates ${\dot m}_0$ and ${\dot m}_0$v$_0$ were
expressed as a function of the velocity of the shock and the jet radius. Here, 
we keep ${\dot m}_0$ and v$_0$ as explicit parameters, so that we can consider a moving surrounding disk wind of arbitrary velocity v$_w$,
and an arbitrarily small $r_j$}.

\subsection{Shape of the bow shock shell}

For a disk-wind with position-independent density $\rho_w$ and velocity v$_w$, the
integrals in \BT{Eqs.}~(\ref{mass}-\ref{xmom}) can be trivially performed, and from
the ratio of \BT{Eqs.}~(\ref{rmom}-\ref{xmom}) one obtains the differential
equation of $r_b(x)$ :
\begin{equation}
\frac{dr_b}{dx} 
=\frac{{\dot m}_0 \textnormal{v}_0}{\pi\rho_w(r_b^2-r_j^2)(\textnormal{v}_j-\textnormal{v}_w)^2}\,,
\label{drdx}
\end{equation}
which can be integrated to obtain the shape of the thin shell bow shock as a function of $x$ \SC{in the IWS reference frame}:
\begin{equation}
r_b(r_b^2-3r_j^2)+2r_j^3=L_0^2x\,,
\label{rx}
\end{equation}
where we defined the characteristic scale\footnote{
\SC{Noting that} $L_0$ is the radius where the swept-up \textit{x}-momentum is equal to 3 
times the injected $r$-momentum ${\dot m}_0 \textnormal{v}_0$, \BT{Eq.}~(\ref{rx}) is equivalent to \BT{Eq.}~(22) in 
\citet{2001ApJ...557..443O}.}
\begin{equation}
L_0\equiv \sqrt{\frac{3{\dot m}_0 \textnormal{v}_0}{\pi \rho_w(\textnormal{v}_j-\textnormal{v}_w)^2}}\,.
\label{l0}
\end{equation}

Clearly, as the thin shell bow shock began to grow at $t=0$, the solution
given by \BT{Eq.}~(\ref{rx}) \SC{must} terminate at a finite maximum radius $r_{b,f}$ (see \BT{Fig.}~\ref{fig:shockframe}). 
\SC{The growth of this outer radius with time} 
can be calculated combining \BT{Eqs.}~(\ref{mass}-\ref{rmom}) to
obtain:
\begin{equation}
\frac{dr_{b,f}}{dt}=\textnormal{v}_r=\frac{{\dot m}_0 \textnormal{v}_0}{{\dot m}_0+\pi\rho_w(r_{b,f}^2-r_j^2)(\textnormal{v}_j-\textnormal{v}_w)}\,,
\label{drfdt}
\end{equation}
which can be integrated with the boundary condition $r_{b,f}(t=0)=r_j$ to
obtain \SC{$r_{b,f}(t)$ at the current time $t$}:
\begin{equation}
\frac{1}{\gamma L_0^2}\left[r_{b,f}^3-r_j^3+3r_j^2(r_j-r_{b,f})\right]+r_{b,f}-r_j=\textnormal{v}_0t\,,
\label{rf}
\end{equation}
with
\begin{equation}
\gamma\equiv \frac{\textnormal{v}_j-\textnormal{v}_w}{\textnormal{v}_0}\,.
\label{beta}
\end{equation}

Now, in order to obtain the shape of the bowshock shell in the \SC{source frame} $(z,r)$  (see \BT{Fig.}~\ref{fig:restframe}, cyan curve),
when the IWS is located at distance $z_j$ from the source, we simply 
insert the relation $x = (z_j-z_b)$  into \BT{Eq.}~(\ref{rx}) and $t = t_j\equiv z_j/$v$_j$ in \BT{Eq.}~(\ref{rf}). 
In the \textit{``narrow jet'} limit where $r_j \to 0$, the thin shell bow shock has the \SC{simple cubic} shape
given by equation:
\begin{equation}
\frac{r_b}{L_0}=\left(\frac{z_j-z_b}{L_0}\right)^{1/3}\,,
\label{rz}
\end{equation}
ending at the maximum ``outer edge" radius $r_{b,f}$ 
(see the cyan dot in \BT{Fig.}~\ref{fig:restframe}) given by Eq.~\ref{rf} evaluated at $t=t_j$:
\begin{equation}
\frac{1}{\gamma}\left(\frac{r_{b,f}}{L_0}\right)^3+\frac{r_{b,f}}{L_0}=\left(\frac{\textnormal{v}_0}{L_0}\right)\,t_j = \left(\frac{\textnormal{v}_0}{L_0}\right)\left(\frac{z_j}{\textnormal{v}_j}\right).
\label{rfz}
\end{equation}
For a \textit{"wide jet''} where $r_j$ is no longer negligible, the corresponding equations can also straightforwardly be obtained
from \BT{Eqs.}~(\ref{rx}) and (\ref{rf}), and are given in appendix \ref{appendixA}. In the following, we will consider the
\textit{"narrow jet"} regime, which leads to simpler equations. 

\subsection{The post-bow shock cavity}

Let us now consider the trajectory \SC{$r_f(z_f)$ described in the $z,r$ plane by} the outer edge of the thin
shell bow shock \SC{at earlier times, when the IWS travelled from its formation point $z=0$ at $t=0$ 
to its current location $z_j$ at time $t_j$}.
This trajectory will define the shape of the volume 
swept out by the travelling and expanding bowshock into the slower disk wind
(see Fig.~\ref{fig:restframe}, dashed black line).

At an earlier time $t_f$  ($0\leq t_f\leq t_j$), the bowshock terminated at an outer radius $r_f \le r_{b,f}$ 
given by Eq.~\ref{rfz} with $t_j=t_f$:
\begin{equation}
\frac{r_f^3}{\gamma L_0^2}+r_f=\textnormal{v}_0t_f.
\label{rftf}
\end{equation}
The distance $z_f$ from the source where this radius $r_f$ \SC{was reached}
is obtained from \BT{Eq.}~(\ref{rz}) by setting $z_b = z_f$, $r_b = r_f$ and $z_j=$v$_j \, t_f$.
\begin{equation}
\frac{r_f^3}{L_0^2} = \textnormal{v}_j t_f - z_f.
\label{zftf}
\end{equation}

Combining \BT{Eqs.}~(\ref{rftf}-\ref{zftf}) to eliminate $t_f$, and recalling 
that $\gamma = (\textnormal{v}_j-\textnormal{v}_w/){\textnormal{v}_0}$\, we then obtain
the shape $r_f(z_f)$ of the cavity swept by the (growing) edge of the bow shock wing associated
with the travelling internal working surface (see dashed black curve in Fig.~\ref{fig:restframe}) :
\begin{equation}
\frac{\textnormal{v}_w}{\textnormal{v}_j-\textnormal{v}_w}\left(\frac{r_f}{L_0}\right)^3+\frac{\textnormal{v}_j}{\textnormal{v}_0}\left(\frac{r_f}{L_0}\right)
= \frac{z_f}{L_0}\,.
\label{zfrf2}
\end{equation}

\subsection{Refilling of the cavity by the disk-wind}

Of course, as soon as the bow shock wing has passed by, the disk wind (travelling in the $z$-direction
at a velocity v$_w$, see \BT{Fig.}~\ref{fig:restframe}) immediately starts to refill the swept-up cavity. 
For a given radius $r_f(z_f)$ along the boundary of the swept-up volume, the refilling by the disk-wind will thus start at the 
time $t_f$ (given by Equ.~\ref{zftf}) when the bowshock edge reached this position; at 
the present time $t_j$ the disk wind will have refilled a region of length $(t_j-t_f)$v$_w$ along the $z$-axis.
The boundary between the \SC{wind-refilled} region and the emptied cavity thus
has a locus $z_c(r_c)$ (see the cyan dash-dotted line in Fig.~\ref{fig:restframe}) given by:
\begin{equation}
z_c=z_f+(t_j-t_f)\textnormal{v}_w=\gamma r_c+\textnormal{v}_w t_j\,,
\label{zc}
\end{equation}
where for the second equality we have used \BT{Eqs.}~(\ref{rftf}-\ref{zftf}) and set $r_c=r_f$.

Therefore, the slower disk wind refills the cavity \SC{swept} by the bow shock except
for an inner, conical ``hole'' with half-opening angle $\alpha=\arctan \gamma^{-1}$ =
arctan[v$_0$ / (v$_j$-v$_w$)]. 
The conical cavity is attached to the wings of the bow shock at $(z_{bf},r_{bf})$, 
and its vertex along the jet axis is located at a distance from the source $z_a$= v$_w\,t_j$ = $z_j$ (v$_w$/v$_j$)
(see \BT{Eq.}~\ref{zc} with $r_c=0$ and cyan asterisk in Fig.~\ref{fig:restframe}).

\begin{figure*}[!t]
\centering
\includegraphics[width=17cm]{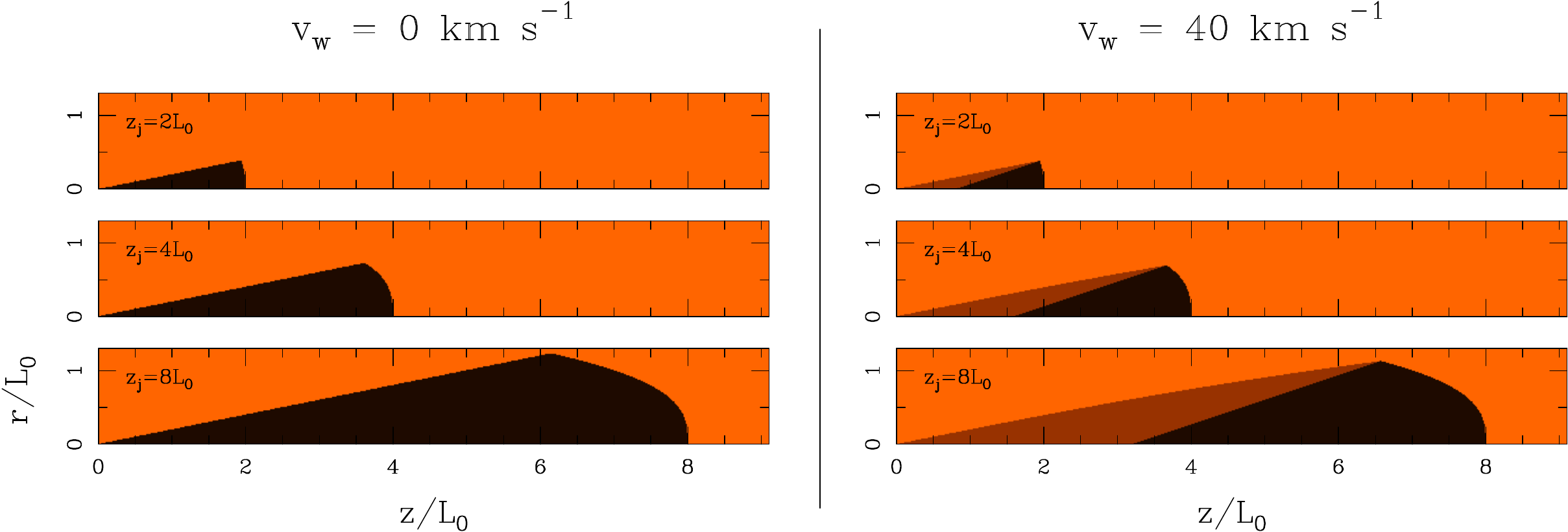}
\centering
\caption{The time evolution of the bow shock + cavity flow predicted by
the analytic model for two choices of the wind velocity: v$_w=0$ (left 
frame) and v$_w=0.4$v$_j$  (right frame).  The dark region is
the empty part of the cavity (swept-up in the thin shell bow shock) and
the brown region is the part of the cavity that has been refilled
by the disk wind (this region being of course absent in the v$_w=0$ model
of the left frame). 
For both models, we show snapshots 
corresponding to $t=2L_0/$v$_j$, $4L_0/$v$_j$ and $8L_0/$v$_j$), which result
in positions $z_j=2L_0$, $4L_0$ and $8L_0$ for the working surface (see the
labels on the top left of each frame). Both models have v$_0=0.2 $v$_j$.}
\label{fig:analytic-timevol}
\end{figure*}

\BT{Fig.}~\ref{fig:analytic-timevol} shows the analytical flow configurations obtained at three different evolutionary times
(corresponding to $t=2L_0/$v$_j$, $4L_0/$v$_j$ and $8L_0/$v$_j$), and for
two choices of the wind velocity (v$_w=0$ and v$_w=0.4$v$_j$). In the two models,
we have set v$_0=0.2$v$_j$. The model with v$_w=0$ (\BT{left} frames of \BT{Fig.~\ref{fig:analytic-timevol}})
produces a cavity which does not fill up. For v$_w=0.4$v$_j$  (right frames
of \BT{Fig.~\ref{fig:analytic-timevol}}), the bow shock has a more stubby shape compared to the v$_w=0$
bow shock ($L_0$ is larger) and the cavity which it leaves behind is partially refilled by
the disk wind (brown region).

\subsection{Kinematics along the shell}

From \BT{Eqs.}~(\ref{mass}-\ref{xmom}), it is straightforward to show that for a narrow
jet surrounded by a homogeneous disk-wind, the radial and axial velocities 
\SC{(in the source rest frame) 
of the well-mixed thin shell material as a function of cylindrical radius $r_b$ are }:
\begin{equation}
\textnormal{v}_r= \textnormal{v}_0 \left(1+\frac{3r_b^2}{\gamma L_0^2}\right)^{-1}\,,
\label{vrv0}
\end{equation}
\begin{equation}
\textnormal{v}_z=\textnormal{v}_w+(\textnormal{v}_j -\textnormal{v}_w) \left(1+\frac{3r_b^2}{\gamma L_0^2}\right)^{-1}\,,
\label{vzv0}
\end{equation}
where for the second equation we have also considered that v$_z=$v$_j-$v$_x$ (see
\BT{Figs.} \ref{fig:shockframe} and \ref{fig:restframe}). In evaluating the radial velocities, one should keep in
mind that the radius $r_b$ is always smaller than the $r_{b,f}$ value given by
\BT{Eq.}~(\ref{rfz}).

As expected, \SC{we find the following asymptotic limits }:
\begin{itemize}
\item $\textnormal{v}_r$ 
has an initial value v$_0$ for $r_b\to 0$ (i.e., as it
leaves the jet working surface) and goes to 0 at large radii (as the radial
momentum of the thin shell bow shock is shared with an increasing mass of
disk wind),
\item $\textnormal{v}_z$ 
has an initial value v$_j$ when it leaves the working surface ($r_b \rightarrow 0$),
and for large radii tends to the disk wind velocity v$_w$.
\end{itemize}

\BT{Eqs.}~(\ref{vrv0}-\ref{vzv0}) give the velocity of the well-mixed material
within the thin shell bow shock. These velocities correspond to the Doppler
velocities observed in an astronomical observation provided that the emission
does indeed come from fully-mixed material.

Another extreme limit is if the emission is actually \SC{dominated by} the gas that has
just gone through the bow shock and which is not yet mixed with the thin shell
flow material. In this case, the axial and radial velocities of the emitting
material would correspond to the velocity directly behind a highly compressive
radiative shock. For such a shock, the velocity of the post shock flow
(measured in the reference system moving with the bow shock)
is basically equal to the \SC{projection of the incoming flow velocity} parallel to the shock front. 
It is straightforward
to show that in this case the immediate post shock radial and axial velocities of the emitting
material \SC{in the source rest frame} are given - in the \textit{"narrow jet"} limit - by:
\begin{equation}
\textnormal{v}_{r,ps}= (\textnormal{v}_j-\textnormal{v}_w) \frac{3 (r_b/L_0)^2}{1+9  (r_b/L_0)^4},
\label{vrps}
\end{equation}
\begin{equation}
\textnormal{v}_{z,ps}= \frac{\textnormal{v}_j+9~\textnormal{v}_w~(r_b/L_0)^4}{1+9(r_b/L_0)^4}.
\label{vzps}
\end{equation}
We note that while v$_{z,ps}$ has the same asymptotic limits as v$_z$ in the full mixing case (see \BT{Eqs.}~\ref{vzv0} and \ref{vzps}), the radial post-shock velocity v$_{r,ps}$ tends to zero both for $r_b\to 0$ 
and for $r_b\to \infty$ (see \BT{Eqs.}~\ref{vrps}), reaching a maximum value of (v$_j-$v$_w)/2$ for a radius equal to $L_0 /\sqrt{3}$. This peak value for
the radial velocity is a general result of bow shock kinematics, valid regardless
of the shape of the bow shock, which was first derived by \citet{1987ApJ...316..323H}.


By combining \BT{Eqs.}~(\ref{vrv0}-\ref{vzps}) with \BT{Eq.}~(\ref{rz}) it is
straightforward to obtain the axial and radial velocities as a function of
the distance $z$ along the symmetry axis. Examples of these dependencies are
shown in the following section.

\subsection{A dimensional example}

\BT{We now consider a particular model of an internal working surface
moving at a velocity v$_j=100$~km~s$^{-1}$, located at $z_j=10^{16}$~cm along the $z$-axis and ejecting side way material at a rate ${\dot m}_0=10^{-8}$M$_\odot$yr$^{-1}$ with a lateral ejection velocity v$_0=10$~km~s$^{-1}$.} 

For the surrounding disk wind, we assume a number density of atomic nuclei
$n_w$=$\rho_w/1.4m_H$ =$10^4$~cm$^{-3}$ and
velocities v$_w=0$ and v$_w=40$~km~s$^{-1}$. With these parameters,
we obtain $L_0=5.2\times 10^{14}$~cm (for v$_w=0$) and $L_0=8.7\times 10^{14}$~cm 
(for v$_w=40$~km~s$^{-1}$). Note that ${\dot m}_0$ and $\rho_w$ are only involved in the shape and kinematic equations through
$L_0 \propto ({\dot m}_0/\rho_w)^{1/2}$, so that only their ratio actually matters in defining the flow properties.

\begin{figure}[!h]
\centering
\includegraphics[width=8cm]{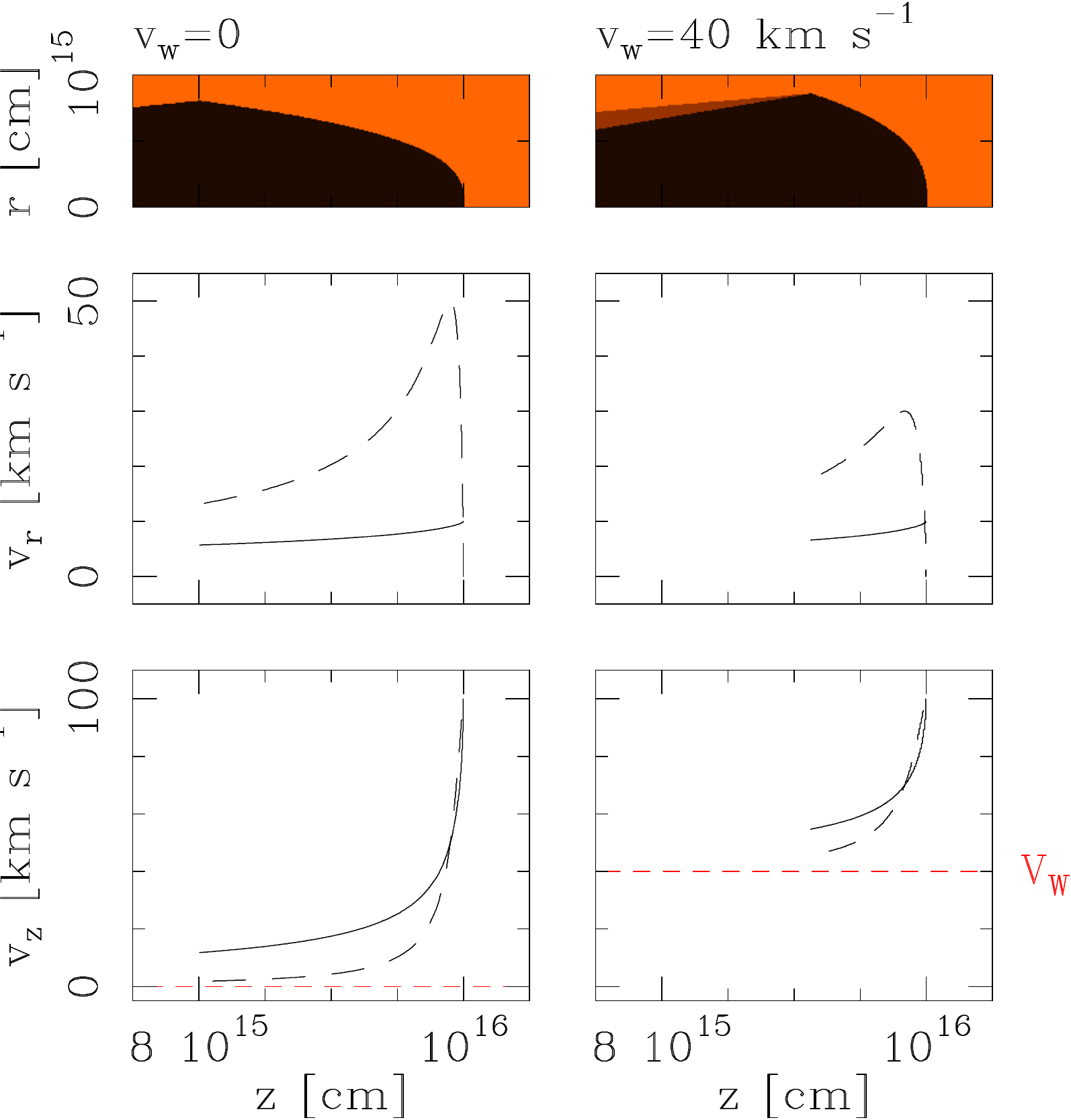}
\caption{Shape of the bow shock and the cavity (top), radial
velocities v$_r$ (center) and axial velocities v$_z$ (bottom) for the
two models discussed in the text. The solid curves show the
velocities of the well-mixed material within the thin shell flow,
and the dashed curves show the immediate
post-bow shock velocities. The dotted red line shows v$_z=$ v$_w$.}
\label{fig4}
\end{figure}

For these two working surfaces, we obtain the shapes, and the radial and axial velocities
(as a function of $z$) shown in \BT{Fig.}~\ref{fig4}. From this figure, it is clear that for the v$_w=40$~km~s$^{-1}$ model
we obtain a flatter working surface than for the v$_w=0$ case, \SC{because $L_0$ is larger.}

The velocities of the fully mixed thin shell material (shown with solid lines
in \BT{Fig.}~\ref{fig4}) have the following behaviors:
\begin{itemize}
\item v$_r$ has a value of v$_0=10$~km~s$^{-1}$ at
$z=z_j$, and monotonically decreases toward  (but not reaching) \BT{zero} for decreasing values of $z$,
\item v$_z$  (lower plots of \BT{Fig.}~\ref{fig4})
has a value of v$_j=100$~km~s$^{-1}$ at
$z_j$, and monotonically decreases for lower values of $z$,
towards (but not reaching) an asymptotic limit of v$_w$.
\end{itemize}

 The immediate post-bow shock velocities (shown with dashed lines in \BT{Fig.}~\ref{fig4}) have the following behaviors:
\begin{itemize}
\item v$_{r,ps}$ (central plots of \BT{Fig.}~\ref{fig4}) is \BT{zero} at the apex of the bow shock surface at $z=z_j$, and rapidly grows to a 
maximum value of
50~km~s$^{-1}$ (for v$_w=0$) and 30~km~s$^{-1}$ (for v$_w=40$~km~s$^{-1}$), 
this value corresponding to (v$_j-$v$_w$)$/2$, as
discussed in the previous section is reached at $z=z_j-L_0/3\sqrt{3}$. 
The radial velocity then decreases again at smaller $z$ until the end of the bowshock wings,
\item the axial velocity v$_{z,ps}$ has the same qualitative behavior
as the well-mixed v$_z$ (see above), but with
a different functional form that approaches faster its limit v$_w$ in the bowshock wings.
\end{itemize}

We expect that in reality, due to incomplete mixing, the emitting material will have axial
and radial velocities between the fully-mixed layer and immediate post-bow
shock velocities shown in \BT{Fig.}~\ref{fig4}. The difference between
these two velocities is particularly important for the
radial component of the velocity of the emitting material.

\subsection{Successive bow shocks}

In the previous section, we assumed that the bow shock associated with an internal
working surface travels through undisturbed disk wind material. However, we saw that the cavity formed
behind it is only partially refilled by fresh disk wind. Therefore, a second bow shock will travel
into a disk wind structure containing an empty, conical cavity left behind by the first bow shock.

\begin{figure*}[!t]
\includegraphics[width=14cm]{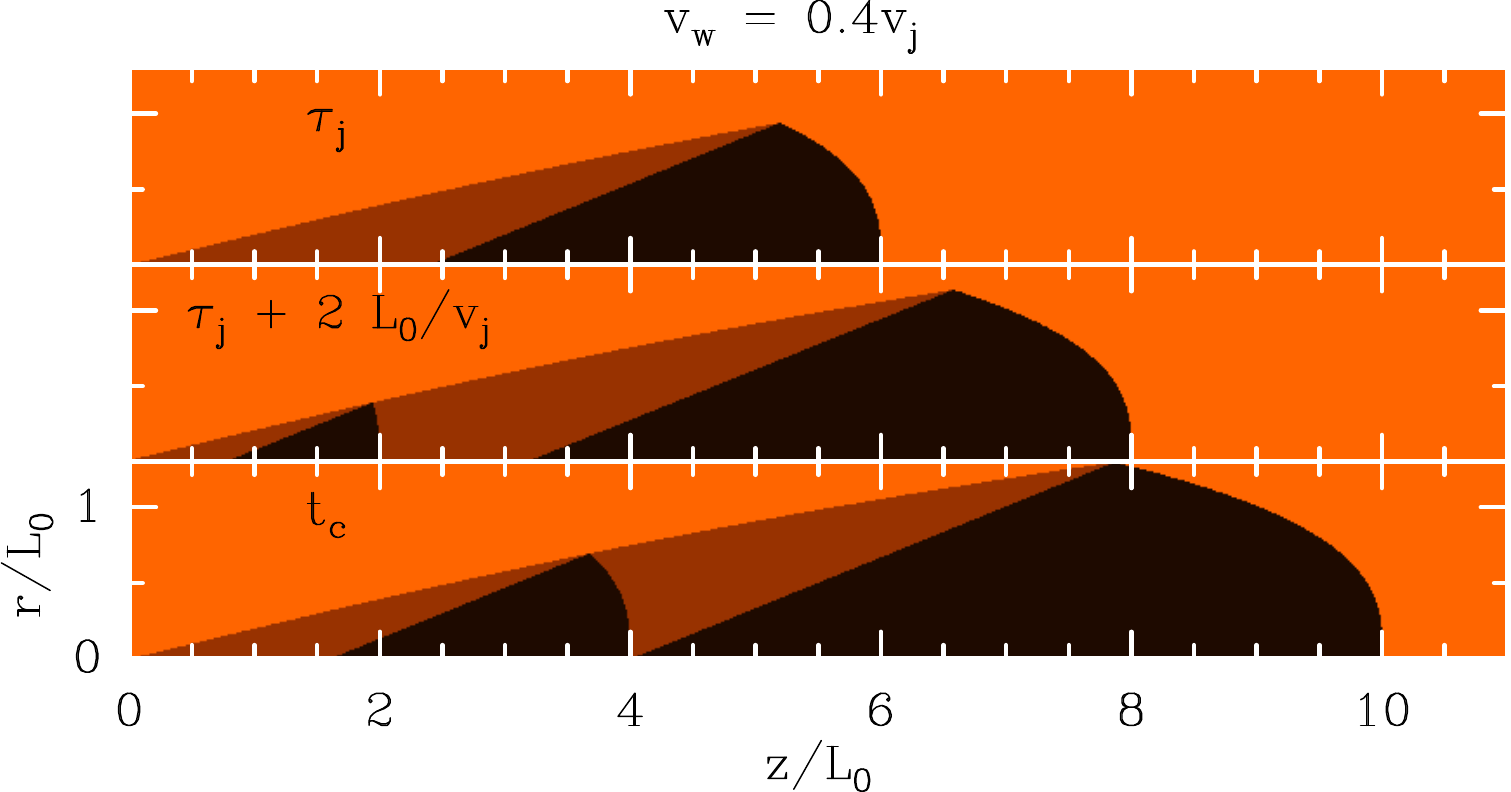}
\centering
\caption{The time evolution of two successive bow shocks and the cavities predicted by the analytical model.
  The first bow shock is ejected at $t=0$, the second shock is formed at time $t= \tau_j$ when the first bow shock
  is at $z=$v$_j \tau_j$ taken here as $6 L_0$ (top). At a time $t=\tau_j+2 L_0/$v$_j$, the second bow shock is still
  propagating in the undisturbed disk wind material (center). At a time $t_c$ the second bow shock catches up the emptied cavity of the first bow shock at its vertex (bottom).}
  \label{fig-twoiws}
\end{figure*}

We now assume that the variable ejection velocity of the jet produces a second working surface 
at $z = 0$ 
at a time $\tau_{j}$, which also travels along the jet axis with
the same velocity v$_j$ as the first working surface.
\BT{Fig.}~\ref{fig-twoiws} illustrates three steps of the propagation of this second working surface.

At $t=\tau_j$ (\BT{Fig.}~\ref{fig-twoiws}, first panel) the first working surface is at a distance $z_{j1} =$v$_j \tau_j$ from the outflow source 
and its cavity is partially filled with fresh disk wind material, while the second working surface 
has not yet expanded.
At a time $\tau_j < t < t_c$ (\BT{Fig.}~\ref{fig-twoiws}, center) the second bow shock travels in unperturbed, pristine disk wind material  
\SC{that refilled the cavity behind the first bowshock};
\SC{hence its shape and kinematics are still given by the same equations derived above for the leading internal working surface,
and any molecules present in the disk-wind can enter the second bowshock.}
\SC{At a time $t = t_c$ (\BT{Fig.}~\ref{fig-twoiws}, bottom panel)  the apex of the second bowshock just catches up with}
the vertex of the conical cavity emptied by the first bow shock and not refilled by the disk wind. 

To obtain the time $t_c$, 
we note that at any time $t$ ($\tau_j < t < t_c$), the position of the apex of the second bow shock is $z_{j2} =$v$_j (t-\tau_j)$ and the position of the vertex of the empty cavity behind the first bow shock is $z_{a1} = $v$_w t$. 
By equating these two quantities we get:
\begin{equation}
  t_c = \frac{\textnormal{v}_j}{\textnormal{v}_j-\textnormal{v}_w} \tau_j.
  \label{criticaltime}
\end{equation}
This interaction occurs at a distance $l_c$ from the source:
\begin{equation}      
  l_c = z_{a1}(t_c) = \textnormal{v}_w t_c =\frac{\textnormal{v}_j}{\textnormal{v}_j/\textnormal{v}_w-1} \tau_j = \frac{\Delta z}{\textnormal{v}_j/\textnormal{v}_w - 1 }
\label{criticaldist}
\end{equation}
where $\Delta z = \tau_j $v$_j$ denotes the distance between two successive IWS.
Unless v$_w$ is very close to v$_j$, we find that $l_c$ is of the order of a few times the typical IWS spacing.

Our model thus predicts that no more pristine unperturbed disk wind material can remain close to the jet axis beyond $z = l_c$. 
When the second IWS reaches $z > l_c$, the central region of its bow shock shell propagates into the emptied
cavity left behind by the previous IWS. This second bow shock shell will in general become less curved than the first one, 
because its central region travels into a low density cavity instead of unperturbed disk wind. 

\section{Numerical simulations}
In the previous section, we proposed a simple analytical "thin-shell" model that describes the morphology and the kinematics of a bow shock produced by a pulsating jet travelling in a surrounding disk wind. We especially show that the disk wind refills up part of the cavity carved by the bow shock, allowing us to observe pristine disk wind close the source. For successive bow shocks, bow shocks travels in an undisturbed disk wind up to a critical distance $l_c$. Above this altitude, bow shock shells interact with each other and analytical models can only be heuristic.

In this section, we present numerical simulations that start with the simple configuration adopted above, first to determine to what extent the analytical model can be used to describe a realistic situation -e.g., with a partial mixing - and secondly to study briefly the long term evolution of the interacting bow shock shells. 

\subsection{Numerical method and setup}

We carry out numerical simulations of a variable ejection jet surrounded by a wide "disk wind" outflow.
We have implemented the new HD numerical code \textit{Coyotl} which solves the "2.5D"
Euler ideal fluid equations in cylindrical coordinates:

\begin{equation}
\frac{\partial \textbf{U} }{\partial t} + \frac{1}{r} \frac{\partial r \textbf{F}}{\partial r} + \frac{\partial }{\partial z} \textbf{G} = \textbf{S}, 
\label{euler2D1}
\end{equation}
where $\textbf{U}$ is the vector of conserved quantities
\begin{equation}
\textbf{U} = (\rho, \rho \textnormal{v}_r, \rho \textnormal{v}_z, e, n_i)
\label{euler2D2}
\end{equation}
with fluxes in the $r$- and $z$- directions given, respectively, by
\begin{equation}
\textbf{F} = (\rho \textnormal{v}_r, \rho \textnormal{v}_r^2+p, \rho \textnormal{v}_z \textnormal{v}_r, \textnormal{v}_r(e+p), \textnormal{v}_r n_i),
\label{eulerU2D3}
\end{equation}
\begin{equation}
\textbf{G} = (\rho \textnormal{v}_z, \rho \textnormal{v}_r \textnormal{v}_z, \rho \textnormal{v}_z^2 +p, \textnormal{v}_z (e+p), \textnormal{v}_z n_i).
\label{eulerU2D4}
\end{equation}
$n_i$ are passive scalars used to \SC{separate} the jet from the \SC{disk}-wind material in the flow.
Assuming an ideal equation of state, the total energy density $e$ is
\begin{equation}
e = \frac{p}{\rho (\gamma -1)} + \frac{1}{2} \rho (\textnormal{v}_r^2 + \textnormal{v}_z^2).
\label{eulerU2D4}
\end{equation}
and the source term is
\begin{equation}
\textbf{S} = (0, \frac{p}{r}, 0, -\rho^2 \Lambda(T), 0).
\label{eulerU2D4bis}
\end{equation}
where the cooling function $\Lambda(T)$ is the parametrized atomic/ionic cooling term
of \citet{1989ApJ...344..404R}, which approximates the cooling curve of \citet{1976ApJ...204..290R} for temperatures above $10^4$K.

The numerical scheme is based on a second order Godunov method with an HLLC Riemann solver (Toro 1999).
The calculation of the fluxes and data reconstruction uses the second order scheme
described by \citet{1991MNRAS.250..581F}. This algorithm solves Euler equations in a true cylindrical coordinate
system as written in \BT{Eq.}~(\ref{euler2D1}) and calculates the cell gradients through
the center of gravity of the cylindrical cells.

We ran two simulations: a reference simulation called \textit{no-DW} model, with v$_w$ = 0 (i.e., a jet in a stationary
ambient medium) and a simulation with v$_w = 0.4 $v$_j$ called \textit{DW} model. 
To follow the refilling of the cavity close to the source and the interaction between various shells, 
we integrate equations on a $2000$ au $\times~350$ au domain, with a resolution of 1~au per cell.
All  jet and wind parameters except v$_w$ are kept equal between the two simulations, and 
are summarized in Table 1.

\begin{table}
\caption{Model parameters}              
\label{table:1}      
\centering                                      
\begin{tabular}{c c }          
\hline\hline                        
Parameter & Value  \\    
\hline                                   
resolution &  $1$ au per cell  \\
simulation domain $z \times r$ & $2000$ au $ \times 350$ au\\      
\hline 
 Jet  \\
\hline     
average jet velocity, $\textnormal{v}_j$ & $96$ km~s$^{-1}$ \\
variability amplitude, $\delta $v$_j$  & $48$ km~s$^{-1}$ \\
variability period, $\Delta \tau_j$  & 33 yr \\
time of velocity increase $\eta \Delta \tau_j$  & 0.1~$\Delta \tau_j$ \\
jet density & $9 \times 10^{-22}$g~cm$^{-3}$ \\
jet temperature  & $28$K \\
\SC{jet radius} & 20 au \\
\hline 
 Disk wind  \\
\hline   
disk wind velocity, v$_W$ & $0$ (\textit{no-DW} reference model)  \\ 
							 & $0.4$ v$_j$ (\textit{DW} model) \\ 
disk wind density & $3 \times 10^{-23}$g~cm$^{-3}$  \\
disk wind temperature & 800K \\

\hline                                             
\end{tabular}
\end{table}

Our initial conditions have an inner, constant velocity jet filling the $r<r_j$ region \SC{at all $z$}, and the disk wind (or 
external stationary medium) filling the rest of the computational domain. This setup differs from the standard
jet initialization in which \SC{the jet is introduced only in a small region
around $z \simeq 0$} and then propagates through the domain, producing a transient with
a leading jet bow shock that sweeps aside the ambient medium \citep{1990ApJ...360..370B,1993ApJ...410..686D}.
By initializing our simulations with a jet that \SC{already} extends across \SC{the whole range of $z$, we do not 
perturb the surrounding medium with the transient leading jet} bow shock and we can directly follow the interaction
between an IWS and the unperturbed disk wind close to the outflow source.

In order to form internal working surfaces in the jet, we impose a saw-tooth ejection variability with a mean velocity $\textnormal{v}_j$, a velocity
jump $\delta $v$_j$ and a period $\Delta\tau_j$. The ejection velocity is assumed to rise linearly from
$\textnormal{v}_j-\delta $v$_j/2$ to $\textnormal{v}_j+\delta $v$_{j}/{2}$ during a time-lapse
$\eta \Delta \tau_j$ and to linearly decrease over a time $(1-\eta) \Delta \tau_j$ back to a velocity
$\textnormal{v}_j-\delta $v$_j/2$. Using the small amplitude velocity variability
approximation of \citep{1990ApJ...364..601R}, we estimate that \SC{for our chosen parameters in Table~1} this variability will produce an internal
working surface in the jet at time $t_S$=5yr and distance $z_S = {(\textnormal{v}_j-\delta \textnormal{v}_j/2)^2}\eta \Delta \tau_j/{\delta \textnormal{v}_j} = 75$ au from the central source, and that the working surface will travel in the jet at a velocity v$_j$ =  96~km~s$^{-1}$.

We adopt a jet radius $r_j = 20$~au consistent with the width of the HH~212 molecular jet obtained from
VLBI measurements by \citet{1998ApJ...507L..79C} and with the widths \SC{of atomic jets} estimated by \citet{2000A&A...357L..61D} close to
the source. The density contrast between the jet and the wind is chosen to be sufficiently high ($\rho_j/\rho_w$ = 29) to produce wide bow shock shell. Temperatures are chosen to insure a transverse pressure equilibrium between the jet and the wind. Note that simulations are not very sensitive to the jet temperature since strong internal shocks cooled down by atomic lines set the temperature of the IWS at T$\sim10^4$ K.

The boundary conditions are reflecting on the symmetry axis ($r=0$) and outflowing in the outer radial and
axial cells. On the $z=0$ boundary, we introduce the jet by imposing fixed constant physical conditions for $r<r_j$.
For $r>r_j$ we impose either the disk wind physical conditions, or a reflecting condition (for the reference simulation
with v$_w$ = 0). In order to avoid numerical problems due to the $z$-velocity shear \SC{between the jet and the surrounding 
disk-wind} we put a velocity gradient on \BT{three} cells (i.e., 3~au) at the outer edge of the jet inflow. 

\subsection{Single bow shock propagation}

\begin{figure*}[!h]
\includegraphics[width=15cm]{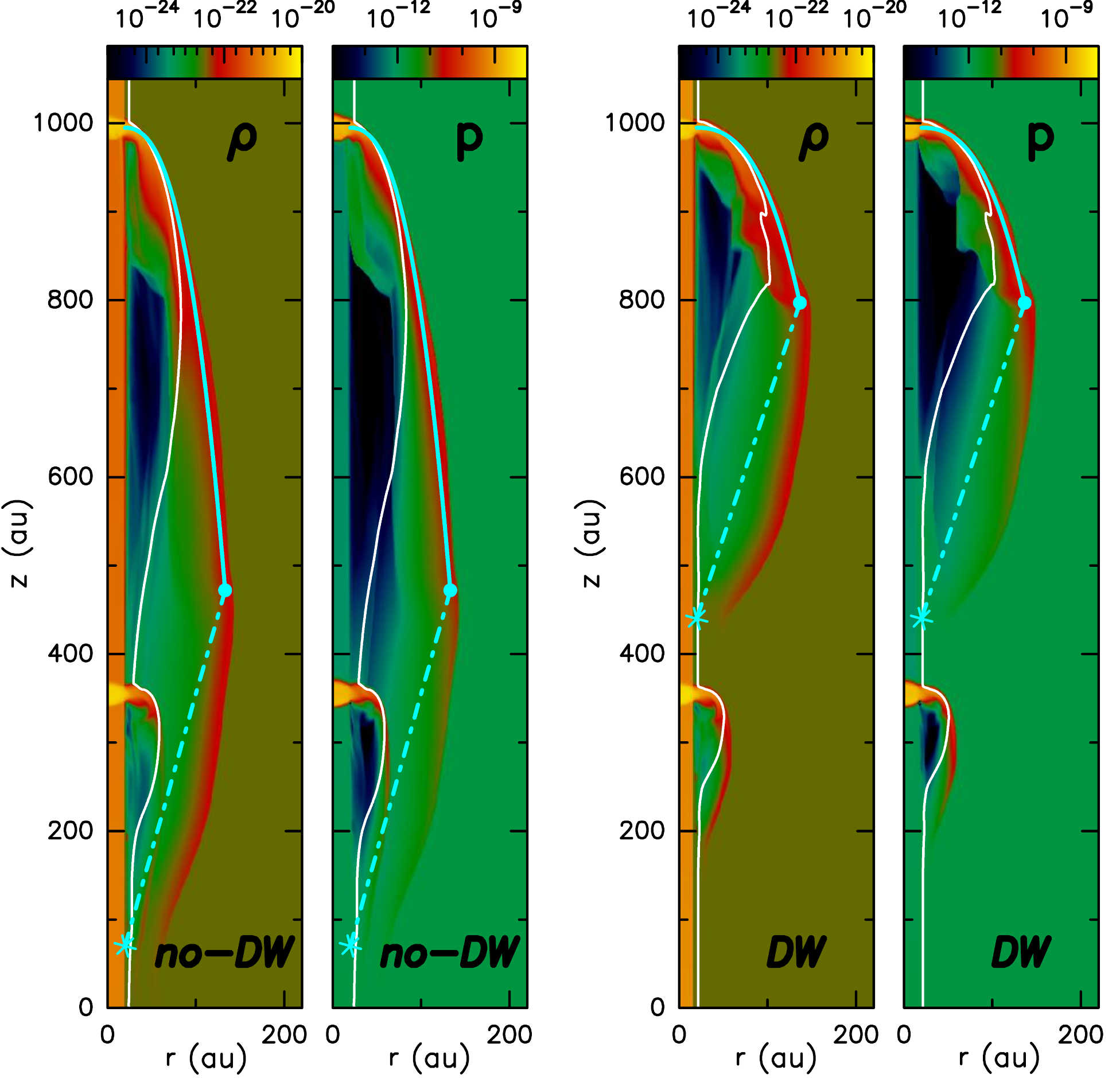}
\centering
\caption{
  Maps of density and pressure for the reference no-DW simulation with v$_w$ = 0 (left) and 
  the  \textit{DW} simulation with v$_w$ = 0.4v$_j$ (right) at a time $t=48$yr. 
  Color scales on top are in g cm$^{-3}$ for density and in dyn cm$^{-2}$  for pressure.
  White contours show the locus of 50\% mixing ratio between jet and disk-wind/ambient material.
  The cyan curve shows a fit (to the numerical results) by the analytic shell shape in \BT{Eq.}~\ref{rfz}, with $L_0 = 65$ au (left) and
  $108$ au (right). 
  The cyan dot indicates the \SC{maximum radius} of 
  the shell, 
 the cyan asterix indicates the predicted vertex of the empty conical cavity left behind the shell, 
 and the cyan dash-dotted line is the analytical predicted boundary between the emptied cavity  and 
  the region refilled from below by fresh disk wind (see Fig.~\ref{fig:restframe} and \BT{Eq.}~\ref{zc}). }
\label{nodw_simu}
\end{figure*}

\BT{Fig.~\ref{nodw_simu}} shows snapshots of the  \textit{no-DW}  simulation (two frames on the left) and of the  \textit{DW}  simulation (two frames on the right)
after a $t=48$~yr time integration, which is larger than
the ejection variability period of 33~yr (see Table 1). The first internal working surface (IWS) has travelled to a distance
of 995~au from the source, and a second IWS to 355~au. 
In this subsection, we study successively the shape of the first bow shock shell, the refilling of the cavity behind it, and the kinematics of the shell, comparing each of them with our analytical predictions.

\subsubsection{The shape of the bow shock shell}

The cyan curves in Fig. 6 show that the bow shock shells in the two simulations can be well fitted with the cubic analytic solution for the thin-shell (\ref{rx}), with values for the characteristic scale $L_0 = 65$~au for the reference \textit{no-DW} simulation and $L_0= 108$~au for the \textit{DW} simulation. 
In the simulations, the sideways ejection velocity v$_0$ and mass-flux  $\dot{m}_0$ (see Section 2)  are a result of the IWS shock configuration, which compresses
the jet material and ejects it sideways.
Since the two simulations only differ in the presence or lack of a surrounding disk wind, the jet IWS in the two simulations have similar characteristics. We then expect that $\dot{m}_0$v$_0$ is the same, and that $L_0$ should vary with the wind velocity as $L_0 \propto ($v$_j-$v$_W)^{-1}$ (see \BT{Eq.}~\ref{l0}).  
The values of $L_0$ found above by fitting the shell shape are indeed consistent with this expectation.

\begin{figure}[!h]
\includegraphics[width=8.5cm]{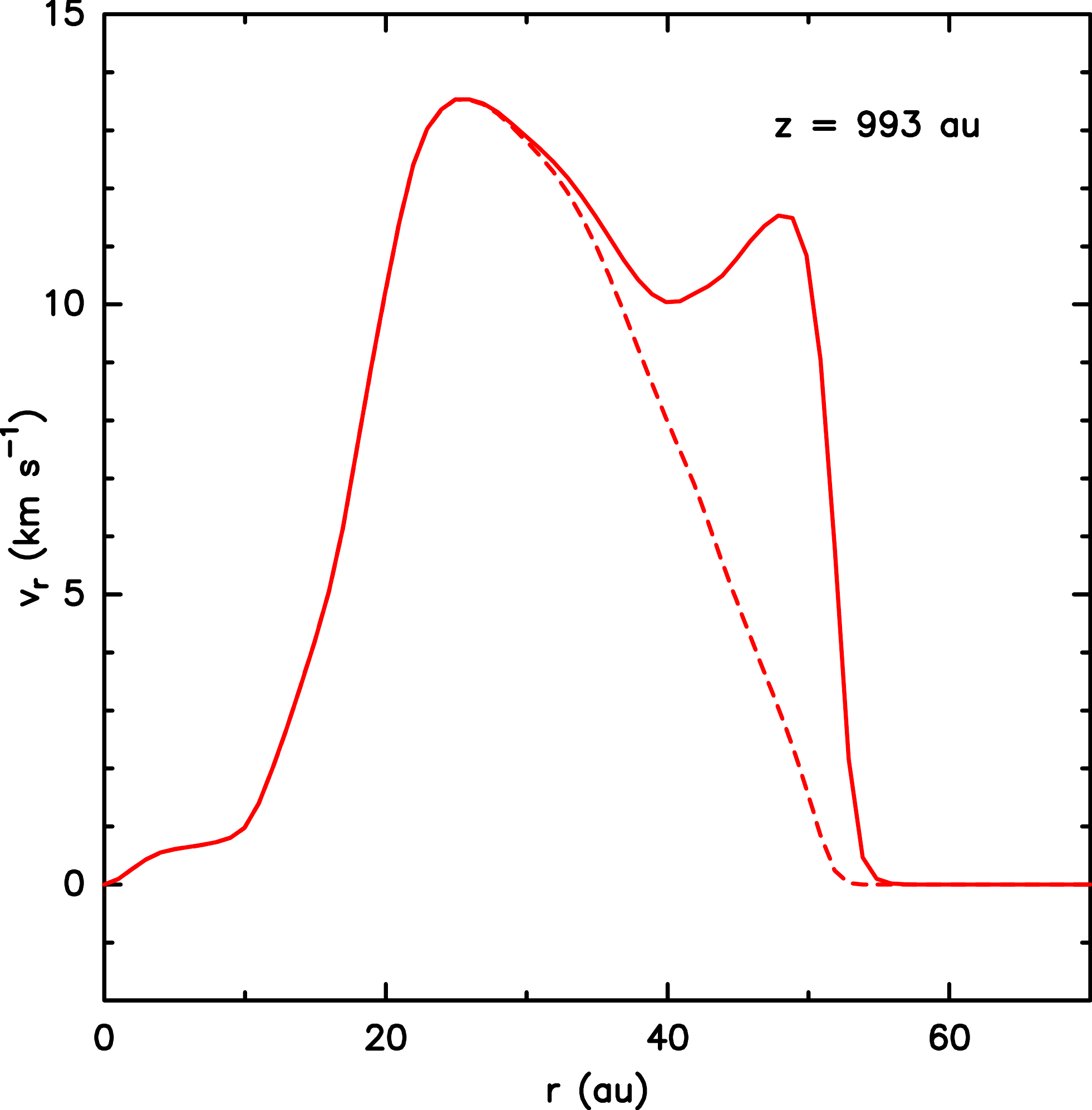}
\caption{Transverse cut across the flow at the IWS location ($z=993$~au) in the \textit{no-DW} time-frame shown in \BT{Fig.} \ref{nodw_simu}. This cut shows the radial velocity
  as a function of distance from the jet axis in solid line. We also plot the radial velocity weighted by the abundance of the jet tracer with a dashed line. The radial velocity first grows outwards, reaches a maximum velocity of $\approx 14$~km~s$^{-1}$
  at a radius of $\sim 25$~au (somewhat larger than the 20~au initial jet radius), and then remains with values $>10$~km~s$^{-1}$ until it drops
sharply to 0 at $r\sim 50$~au. The velocity maximum at $r \sim 25$~au corresponds to the shock against the jet material. The second maximum at r$\sim 50$~au is the shock that propagates in the disk-wind and the zero radial velocity material at larger radii is the undisturbed disk wind.}
\label{v0est}
\end{figure}

The cyan dots on the leading bow shock wings in \BT{Fig.} \ref{nodw_simu} indicate the maximum radius of the bow shock shell as observed in the numerical simulations,  $r_{\rm max} = $ 133 au in the \textit{no-DW} and $r_{\rm max} = $ 137 au in the \textit{DW} cases. Assuming that it corresponds to 
the current position of the edge of the thin shell $(r_{bf},z_{bf})$, as defined in \BT{Fig.}~\ref{fig:restframe},  our analytic model  predicts  that $r_{bf}$ depends on $L_0$ and v$_0$ through \BT{Eq.}~(\ref{rfz}) or (\ref{rf-wide}). With $L_0=65$, $108$~au, we would deduce v$_0 = 27$, $19$ km~s$^{-1}$ for
the \textit{no-DW} and  \textit{DW} simulations, respectively. 

To obtain a direct measurement of v$_0$, 
we plot in \BT{Fig.} \ref{v0est} a transverse cut of the radial velocity v$_r$ at the position of the leading working
surface ($z=993$~au) in the \textit{DW} simulation. Inside the IWS, because of both the adiabatic expansion and the mass flux across the IWS,
$u_r$ increases from zero to a maximum velocity of
$14$ km~s$^{-1}$. This direct measurement of v$_0$
is smaller than the values of 27, 19~km~s$^{-1}$ inferred from the maximum radius of the shell in our simulations using \BT{Eq.}~\ref{rfz} (see above). 
However, we note that taking v$_0$ = 14 km~s$^{-1}$ (its real value), the predicted $r_{bf}$ would be $r_{bf} = 111$~au in the no-DW case and $r_{bf} = 121$~au in the DW case, only slightly
smaller than the $r_{\rm max}$ found in our simulations.

This slight difference in outer radius 
between the analytic model and the numerical simulations could be a result of several effects:
\begin{itemize}
\item The analytic model assumes a working surface with a time-independent, sideways ejection,
  while the numerical simulation has an IWS with time-dependent sideways ejection that depends on the evolution
  of the IWS shocks. The IWS in the simulations produces a higher sideways velocity at early times (v$_0 \sim 18$~km~s$^{-1}$)\footnote{Note that following \citet{2001ApJ...557..443O} the maximum velocity that an atomic gas at T$=10^4$K can reach through adiabatic expansion is $\sqrt{3} c_s = 18$~km~s$^{-1}$, where $c_s$ is the adiabatic sound speed.}, closer to the values deduced from the analytic cavity shapes,
\item in the numerical simulation, the sideways ejection from the IWS is not highly super-sonic. The thermal gas pressure is therefore expected
  to be an additional source of sideways momentum for the shell (an effect not included in our momentum conserving analytic model); \SC{this will act to
  produce a higher ``effective" v$_0$.}
\item similarly, the thermal pressure in the head of the bow shock driven into the surrounding environment will result in
  a sideways push which is not present in the momentum conserving analytic model. 
\item the numerical simulations do not have instant mixing between the sideways ejected jet material and the shocked
  environment (or disk wind), as assumed in the analytic model. Since the immediate post shock velocity in the radial direction is generally greater that the radial mean shell velocity (see example in \BT{Fig.}~\ref{fig4}), the growth rate of the bow shock can be enhanced. \SC{In the reference frame of the IWS (see Fig.~\ref{fig:shockframe}), the non-mixed material will "slide" along the shell surface, extending $r_{b,f}$ to larger values.}
\end{itemize}
\citet{2001ApJ...557..429L} found in their simulations similar disagreements between direct measurements of the sideways momentum ejected by the IWS
and the momentum estimated from the fitted shape of their analytic shell model.

\subsubsection{Cavity refilling}

The asterisk in cyan in each panel of \BT{Fig. \ref{nodw_simu}} indicates the location of the vertex of the emptied cavity as predicted from the analytic model
(see \BT{Fig.}~\ref{fig:restframe}). For the \textit{no-DW} simulation, this point is located at the shock formation position ($z_s=75$~au)
whereas for the \textit{DW} simulation this point is located at $z_a = z_{sf}$ + $\frac{v_w}{v_j} (z_j - z_{sf}) = 440$~au. 
We also plot in cyan dash-dotted \SC{the line connecting this vertex to $r_{\rm max}$,} which traces the 
boundary predicted by the analytical model between the emptied swept-out conical cavity and the unperturbed surrounding medium/refilled disk wind
(see black conical region in Fig.~\ref{fig:restframe}). Three important features can be seen.

\textit{In both numerical simulations, the emptied cavity predicted by the analytical thin-shell model,
i.e. the conical volume inside the dash-dotted cyan line,
is not really empty, but partially filled with a cocoon of low
density and pressure material.} 
No unperturbed ambient gas or disk wind can be left inside this volume (in black in Fig. \ref{fig:analytic-timevol}), which was entirely swept-out by the growing shell 
during the IWS propagation.
Hence this cocoon is made of shocked material that did not fully mix in the shell, and re-expanded in the low-pressure cavity behind it,
 refilling it ``from above''.
The white contour, which denotes the surface of 50\%\  jet/environment mixing fraction (obtained following a passive scalar) 
\SC{shows that} the cocoon is mainly filled with jet material close to the axis, where the shell mass is dominated by 
\SC{gas ejected from the IWS}. 
\SC{Further from the axis and closer to the theoretical boundary (cyan dash-dot line)} 
it is filled by ambient material that was swept up by the bowshock and re-expanded behind it.

\textit{The boundary with unperturbed ambient \BT{or} disk wind material} 
closes back to the axis at the predicted vertex position (see cyan asterisk \BT{Fig.}~\ref{nodw_simu}), 
but is delimited by a weak shock that extends slightly outside from the predicted analytical boundary (dash-dotted cyan line \BT{Fig.}~\ref{nodw_simu}) .
In the \textit{no-DW} model, the analytical boundary 
represents the trajectory of the edge of the bow shock ($z_c = z_f$ in the case v$_w=0$).
Hence, this weak shock is produced by the supersonic motion of the high-pressure edge of the bow shock ($r_{b,f},z_{b,f}$) in the static surrounding medium.
This launches a weak outward shock that repels the boundary of the unperturbed 
ambient 
material slightly outside the predicted cavity boundary (in cyan dash-dotted line). 
In the presence of a supersonic disk wind, the weak shock front is advected away from the source so that it still closes back 
on-axis at the predicted vertex position $z_a$. 
Hence, \BT{Eq.}~(\ref{zc}) gives a strong limit on the boundary between perturbed and unperturbed material. 


\textit{In the presence of a disk-wind, the region between the predicted cavity boundary (cyan dash-dotted line) and the weak shock front outside it
is refilled by fresh disk wind material coming from below.}
\begin{figure}[!h]
\includegraphics[width=9cm]{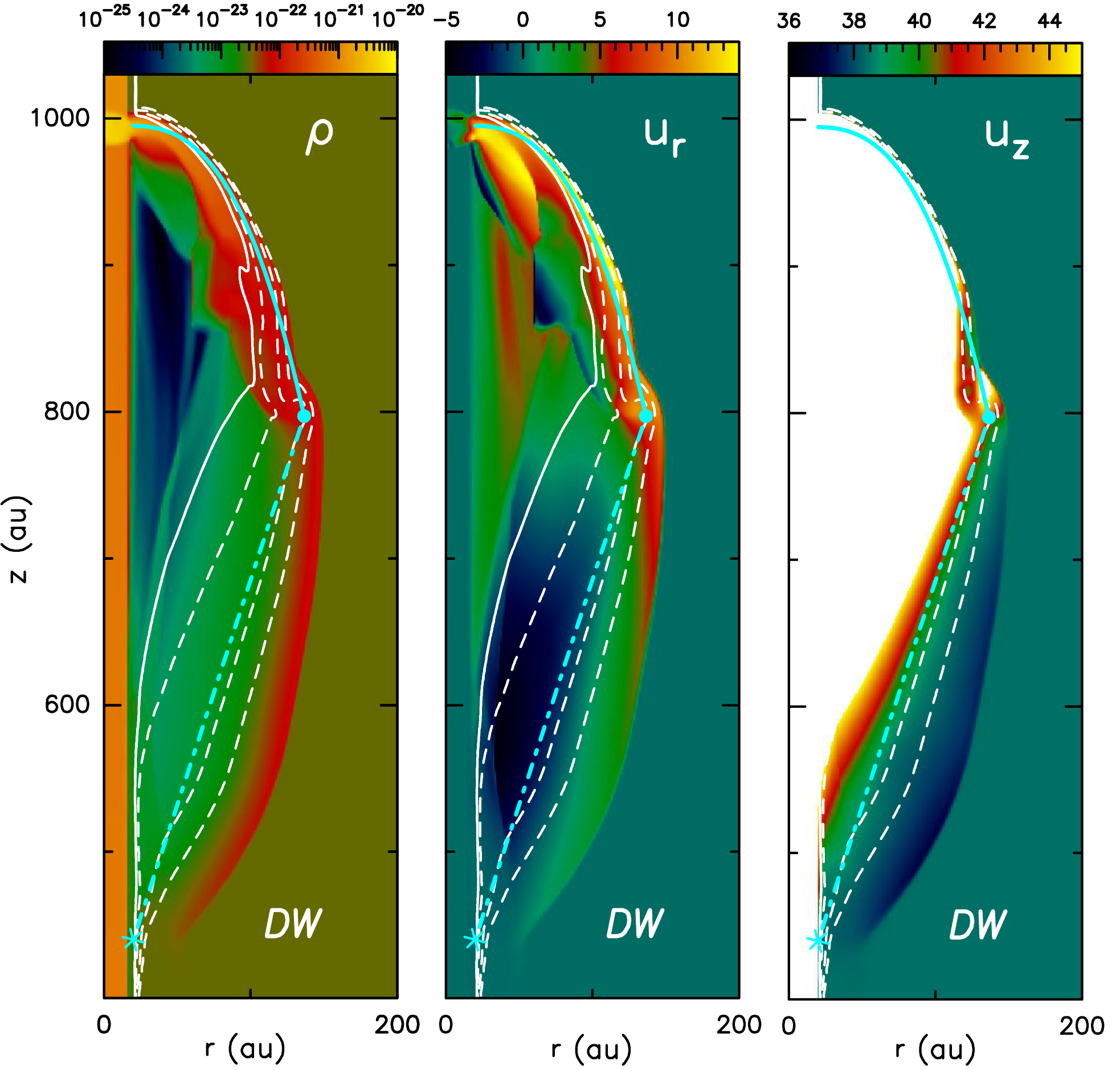}
\caption{Zoom of the leading IWS of the simulation with a surrounding disk wind at time $t=48$~yr. Left: density
stratification (with the logarithmic color scale given by the top bar in g~cm$^{-3}$), center: radial velocity
(with the linear scale of top bar in km~s$^{-1}$) and right: axial velocity structure
(with the linear scale of the top bar in km~s$^{-1}$). The white contours show the surfaces of
50\% (solid line), 10\%, 0.1\% and 0.001\% (outer contour) jet material mixing fractions. 
The cyan asterisk is the predicted vertex of the cavity, the cyan dash-dotted line in the predicted boundary 
of the cavity, and the cyan curve is the fitted shape of the bow shock.}
\label{refilling}
\end{figure}
To analyse this process
we show in Fig. \ref{refilling} density and velocity maps of the region around the leading bow shock of the \textit{DW} simulation.
The dashed white contours show  
10\%, 0.1\% and 0.001\% jet material mixing fractions. 
Material that went through the bowshock and re-expanded in the cocoon has also been partially mixed with jet material. As a consequence, regions where no jet material is observed are regions where the disk wind has refilled the cocoon from below.
The location of the last, outer contour (corresponding to a $10^{-5}$ jet material mixing fraction)
shows that the weak outer shock front propagates into un-mixed, fresh DW material. This material manages to cross the weak 
shock to refill ``from below" the bottom part of the swept-out cavity.

The weak shock provides a slight push outwards to the refilling DW, with radial velocities 
that vary from $+6$~km s$^{-1}$ to $+3$~km s$^{-1}$ along the shock front (middle panel of Fig.~\ref{refilling}), 
similar to the adiabatic sound speed of the disk wind ($c_{s w}\approx2.8~$km~s$^{-1}$). 
The weak shock also reduces the DW inflow velocity v$_z$ to values slightly below v$_w$ = 0.4v$_j$ = 38.4~km~s$^{-1}$ (right panel of Fig.~\ref{refilling}).
However,  refilling remains efficient up to the locus predicted by our analytical model (dash-dotted cyan line), 
as the jet mixing fraction there remains very small ($\simeq$ 0.1\%). 
The presence of the weak shock does not appear to significantly modify the extent of DW refilling compared to analytical expectations.


In summary, we can therefore distinguish in our simulations three refilling regions behind the bowshock:
  \begin{itemize}
  \item a \SC{low density cocoon} trailing the bow shock, that is refilled ``from above" by shell material re-expanding into the emptied cavity.
  This region is mainly composed of jet material close to the apex of the bow shock, and of shocked swept-up disk wind material 
  behind the wings of the bowshock,
  \item an intermediate region (outside the cyan dash-dotted line and inside the weak shock closing the cavity) refilled ``from below''  
  by weakly shocked disk wind material, 
    \item a region upstream of the weak shock closing the cavity, that is refilled by unperturbed fresh disk wind keeping its initial physical conditions.
  \end{itemize}

\subsubsection{Kinematics}

\begin{figure}[!h]
\centering
\includegraphics[width=9.5cm]{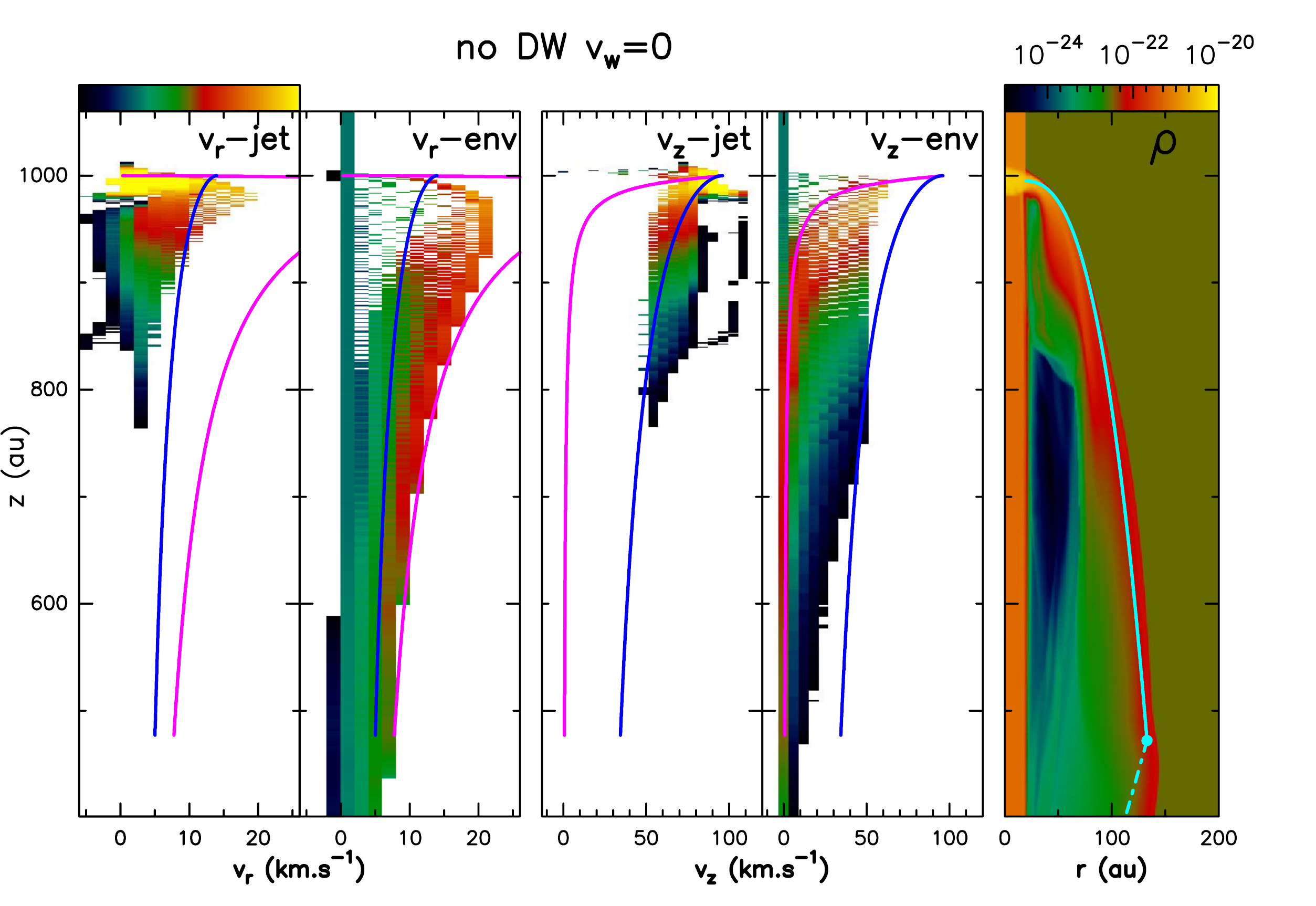}  \\
\includegraphics[width=9.5cm]{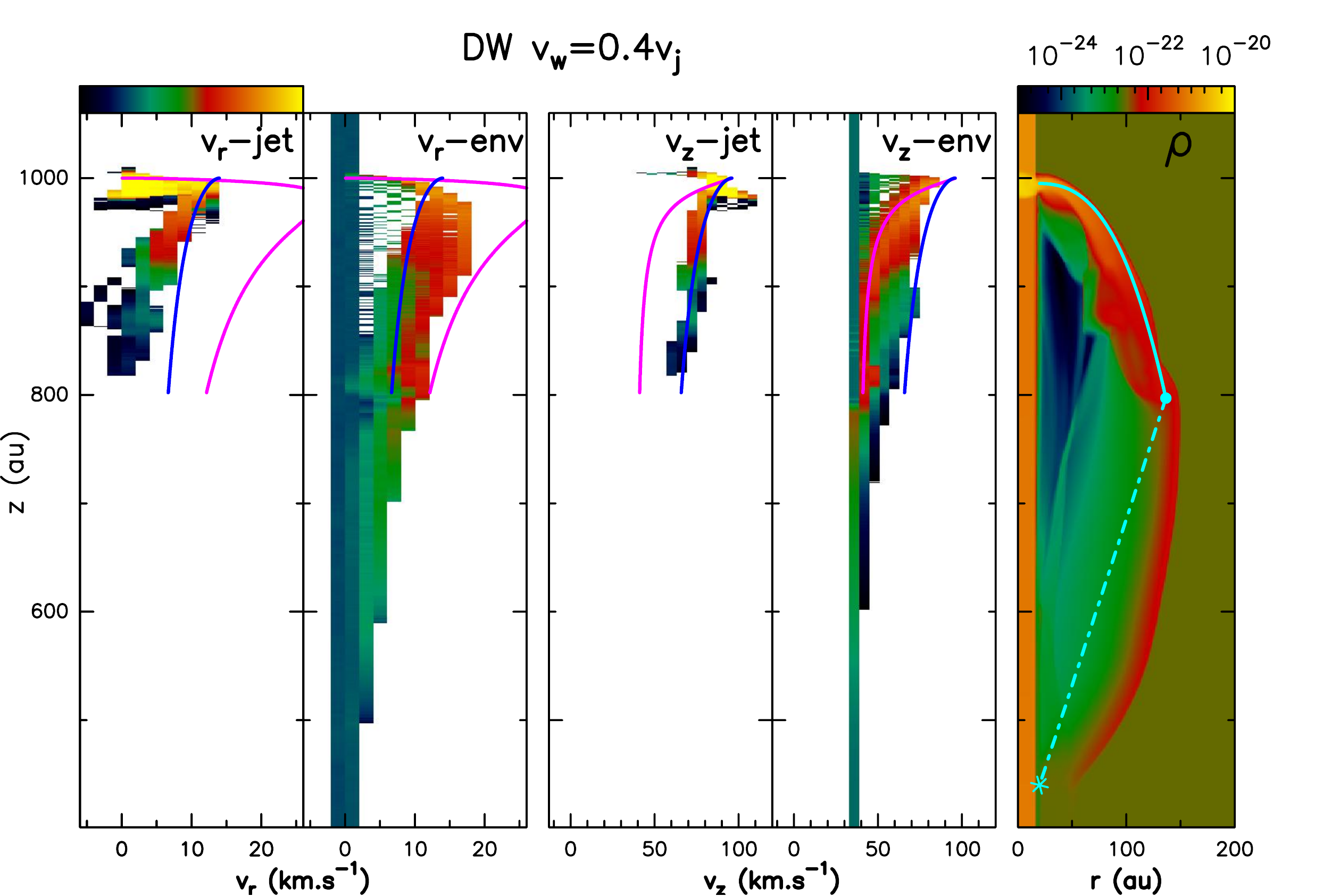}
\caption{Longitudinal position-velocity (PV) diagrams for the \textit{no DW} simulation (top) and the \textit{DW} simulation (bottom).  From left to right: v$_r$ for the jet material
only, v$_r$ for the surrounding material only, v$_z$ for the jet material only, v$_z$ for the surrounding material only, and density stratification. The ordinate of all frames
is position along the outflow axis (in au). The color scale in the PVs is scaled by volume $\times$ cube of pressure so as to be maximum (in red and yellow shades) 
for shocked material in the shell, while the \SC{color scale} for density is in g~cm$^{-3}$. 
Blue curves are predicted velocities in the full mixing hypothesis (\BT{Eqs.}~\ref{vrv0} and \ref{vzv0}), while
magenta curves are the predicted immediate-post shock velocities (\BT{Eqs.}~\ref{vrps} and \ref{vzps}).}
\label{PV}
\end{figure}

We now compare the kinematics in both simulations with our analytical predictions.
\BT{Fig.}~\ref{PV} shows "position-velocity" (PV) diagrams for v$_r$ and v$_z$ as a function of distance $z$ along the flow axis.
In order to enhance the contribution from the material that has just been shocked, each pixel in a snap shot has been weighted by the cube of the pressure $p^3$ times the elementary volume $2\pi r \Delta r \Delta z$.
Using this 
weighting, the maximum intensity (in yellow and orange shades) at each position in the PV diagrams then traces the velocity in the shell.
The separation between material originating mainly from the jet or mainly from the surroundings/disk wind is done using a passive scalar.
The predicted mixed shell velocities (\BT{Eqs.}~\ref{vrv0} and \ref{vzv0}) are shown in blue, and the predicted immediate post-bowshock velocities (\BT{Eqs.}~\ref{vrps} and \ref{vzps}) are shown in magenta. Following the discussion of \BT{Fig. \ref{v0est}}, we take v$_0= 14$ km s$^{-1}$. 

In the v$_r$ PV diagrams of the surrounding
material, the \SC{expansion velocities of shocked material in the shell (orange shading) are always larger than predicted by the (blue) full-mixing curve}
(except very close to the bow shock apex where the shear is maximum). The 
simulation more closely follows the immediate post-bow shock velocity curve (magenta), \SC{indicating that high-pressure shocked material in the shell
is not fully mixed in our simulations.} \SC{Conversely,} the v$_r$ PV diagram of the jet material 
decreases monotonically along the bow shock wing 
with velocities always \SC{slightly smaller than} the full mixing velocity curve (in blue).  

Concerning the velocity along the jet $z$-axis,
the v$_z$ values for jet material lie close to, or slightly above the full mixing curve in blue.
The v$_z$ PV diagrams for the surrounding material 
\SC{generally show smaller v$_z$ than predicted by the full mixing curves.}
\SC{The high-pressure swept-up shell material (in orange)} 
lies close to the immediate post-shock velocities (magenta curve).
 
The relatively small v$_r$ velocities and large v$_z$ velocities 
observed in the jet dominated material indicate that even if the full mixing hypothesis does not hold, the momentum is still conserved: if the velocities of the swept-up surrounding material are greater than expected from the full mixing hypothesis, then the velocities of the jet material (in the IWS rest-frame) must be smaller than the predicted full mixing velocities (and vice-versa). 
 
As predicted, the most striking difference between the disk wind model and the reference no-DW model is
the saturation of the v$_z$ velocity in the bowshock wings \SC{to a non-zero value of} v$_z \approx $v$_w$. Even if this asymptotic limit does
not depend on any mixing, the \SC{incomplete} 
mixing obtained in the simulations produces a \SC{more} rapid convergence to v$_w$
\SC{than predicted in the case of full mixing (blue curve)}.
 
 \subsection{Long-term evolution}

\begin{figure*}[!h]
\includegraphics[width=6cm]{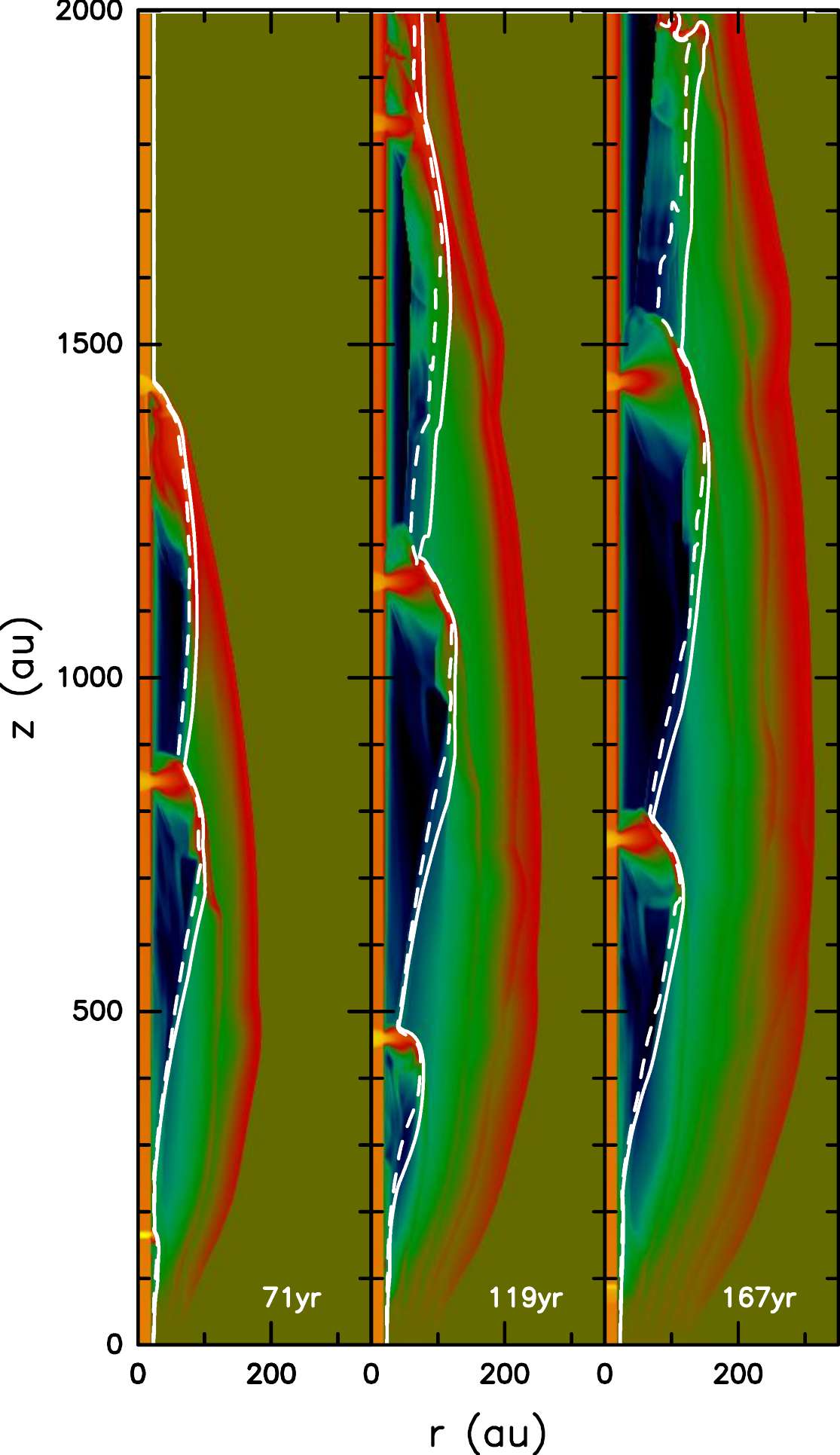}
\includegraphics[width=7.97cm]{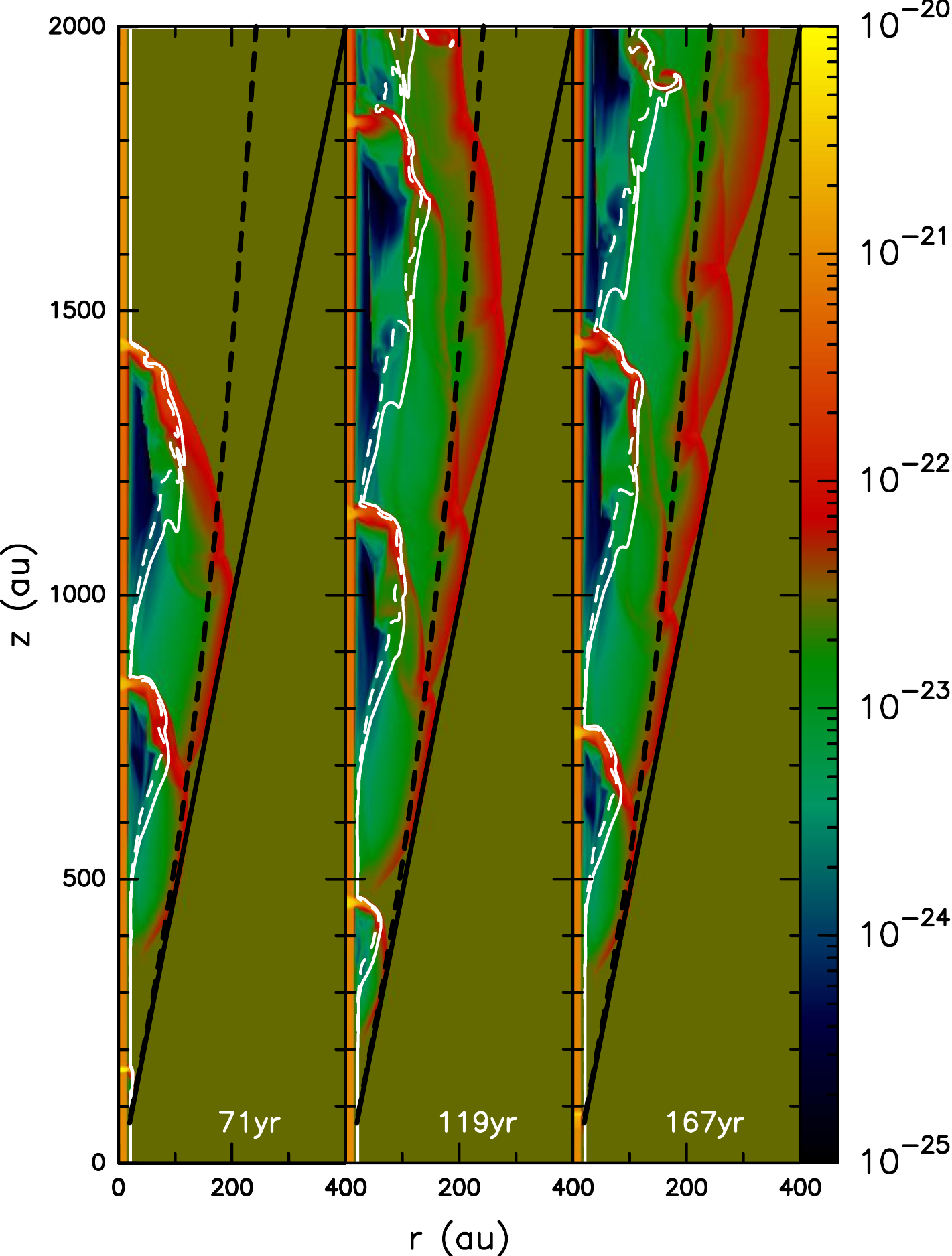}
\centering
\caption{Density maps for the \SC{no-DW} reference simulation (three frames on the left) and the  \SC{DW simulation} 
(three frames on the right) at $t=71$, 119 and 167~yr. The white contours indicate the surface of $50\%$ (solid line) and 90 \% (dashed line) 
jet material mixing fractions. The black lines in the disk-wind simulation show
a cone of $\alpha = 11^\circ$ opening half-angle, which circumscribes the
boundary the region perturbed by the jet and its internal working surfaces. The black dashed lines show the predicted trajectory of the edge of the bow shock (see eq. \ref{zfrf2}). The density color scale is given by the right bar (in g~cm$^{-3}$).}
\label{long_term}
\end{figure*}

\BT{Fig.} \ref{long_term} shows the longer-term evolution of the 
reference no-DW simulation (\BT{three} left frames) and of the disk-wind simulation (\BT{three} right frames)  at times $t=71$, $119$ and $167$~yr.
From this figure,
we see that the morphologies of the regions perturbed by the jet after the passage of several IWS
are very different in the two cases.

In the \textit{no-DW} simulation, the region perturbed by the jet behind the leading bowshock
expands into a roughly
cylindrical shape, which tapers off close to the position of the outflow source (where it becomes a weak, radially expanding shock). This
is the standard shape of the perturbed region in simulations of variable, radiative jets propagating into a uniform static medium,
seen since the early
work of \citet{1993ApJ...413..198S} and \citet{1994ApJ...434..221B}.

In the disk wind simulation, in contrast, the
region perturbed by the jet behind the leading bowshock takes
a conical shape, tapering off at large distances from the outflow
source.  
For the parameters of our DW simulation, the half-opening angle of the perturbed region 
is $\alpha\approx 11^\circ$ (see Fig.~\ref{long_term}). 
This cone is located outside the predicted trajectory of the edge of the bow shock (drawn in black dashed line) given by \BT{Eq.}~(\ref{zfrf2}). This broadening occurs because, as discussed in item 3 of Section 3.2.2, the edge of the bow shock drives a weak outer shock into the undisturbed DW, which propagates away at a speed close to $c_s\sim3.8$ km~s$^{-1}$. Taking into account the advection of the weak shock by the DW, one predicts that this will broaden the disturbed region by an angle  $\beta = \arctan{c_s/\textnormal{v}_w} = 4\degr$, in agreement with the observed cone opening. Obviously, in the no-DW simulations, this weak outer shock travels laterally without being advected, and no limiting cone forms.

In this surprisingly simple configuration
adopted by our jet+disk wind simulation, the overall long-term effect of the disk wind is
to stop the perturbations from travelling beyond this ``opening cone'' of the sideways ejection from the IWS. 

Another important effect of the DW is to push the locus of 50\% ambient material (white contour) 
closer to the jet axis than in the no-DW case, due to the disk wind partial refilling behind each bowshock.
Hence, the first few IWS close to the source can still sweep up (possibly molecular) DW material.
The internal IWS are also more curved than in the no-DW case, where material ejected sideways meets a very low-density cocoon,
producing flat-topped internal bowshocks (see \BT{Fig.} \ref{long_term}).

\section{Summary}

In this paper we have presented a first exploration of an hydrodynamical flow composed by an inner,
variable jet surrounded by a slower, steady, cylindrical disk wind. The jet variability
produces internal working surfaces (IWS) which drive bow shocks into the disk wind, producing
a strong coupling between the two components of the flow.

We have developed a standard thin shell model for the bow shock driven into the disk wind
by a single IWS, for a jet of arbitrarily small radius  (see Section 2), 
deriving the shape of the bow shock and the refilling by the continuing disk wind
of the cavity left behind by the bow shock. The model was extended
to give a qualitative description of the flow resulting from two or more successive IWS bow shocks
plowing through the disk wind (Section 3).

The appropriateness and limitations of the predictions of bow shock shapes and kinematics from this analytic model 
have been checked with axisymmetric numerical simulations:
one of a variable jet+disk wind configuration,
and a second reference simulation with the same variable jet surrounded by a stationary
environment. We compared the analytic model with the numerical simulations, and we found
a relatively good agreement, giving us an understanding of the main features of
the simulated flows. These features are:
\begin{itemize}
\item the bow shocks of the numerical IWS have cubic morphologies which can be reproduced
quite convincingly with the thin-shell, momentum conserving analytic
model (see \BT{Eqs.}~\ref{rz}-\ref{rfz} and Fig. \ref{nodw_simu}),
\item the kinematics in the simulated bow shocks has a behavior
which approximately follows the kinematics predicted from the analytic model
for the fully-mixed layer (for jet-dominated material) or 
the immediate post-bow shock gas (for high-pressure swept-up ambient gas)  (see Figs. 4. and 8).
\item these bow shocks leave behind cavities which are partially refilled
by the slower disk wind (see Figs.~3, 5 and 8).
\item thanks to this refilling, subsequent IWS will propagate into fresh disk wind material 
up to a distance from the source $l_c = {\Delta z}$/(v$_j$/v$_w$-1) (see Fig.~7).
\end{itemize}

The main contribution of this paper is thus to provide a numerically validated, simple analytic model which can
be used to model bow-like shapes of knots observed close to the outflow sources in high velocity,
collimated optical and molecular outflows \citep{2000A&A...356L..41L,2015A&A...581A..85P}.
As shown by our simulations, this shape modeling (in the narrow jet limit) allows one to estimate the sideways ejection velocity 
from the IWS and the length scale of the bowshock. From this, constraints could be inferred on the 
mass-loss rate from the IWS and the surrounding flow properties (see \BT{Eq.}~\ref{l0}).

Another important contribution of this paper is to predict the regions
where a surrounding disk-wind would remain unperturbed.
A quite dramatic result of our jet+disk wind simulation is that the perturbations of the disk wind by the IWS bow shocks 
are confined inside a cone.
Therefore, all of the gas outside this confinement cone is unperturbed disk wind material. Also,
there are pockets of undisturbed disk wind material within this cone, in the refilled region
between the source and the last IWS, and also ahead of the latest
IWS when it is at $z < l_c$ (see the three right hand frames of Fig. 10). 
These are the regions in which one still 
finds a record of the undisturbed characteristics of the disk wind, 
which could be useful for comparisons with disk wind models.

Finally, another result of observational interest is that we identify several 
distinctive signs of a cylindrical DW around a time-variable jet:
(i) bow shocks that close upon the axis at a finite distance from the source (at a fraction v$_w$/v$_j$ of the distance to the bow apex), 
(ii) a non-zero (= v$_w$) asymptotic value of longitudinal velocity in the far bowshock wings,
(iii) internal bowshocks that are curved rather than "flat-topped", 
(iv) a predominance of DW material ahead of the first few IWS, which (if the DW is chemically richer and/or dustier than the jet) 
should produce different emission signatures compared to the more distant IWS.

Extensions of the analytic model to more complex jet+disk wind flows do not
appear very attractive (as, e.g., relaxing the assumption of a cylindrical uniform disk
wind) as quite complex expressions are obtained, and are therefore not straightforwardly
applicable to model observed structures. On the other hand, the numerical simulations
presented here can be extended in many directions:
\begin{itemize}
\item including a more realistic disk wind model (e.g., with a radial dependence of the density and velocity, and a velocity not aligned with
the outflow axis),
\item studying the effect of a non-top hat jet cross section,
\item going from the HD to the MHD equations,
\item including a chemical/ionic network and the associated cooling functions.
\end{itemize}
If future comparisons between jet+disk wind models and observations are sufficiently
promising, the items listed above (as well as other easily imagined possibilities)
will become worthy of exploration.

\begin{acknowledgements}
\BT{We thank the anonymous referee for useful comments.} This work has been done within the LABEX PLAS@PAR project, and received financial state aid managed by the Agence Nationale de la Recherche, as part of the ''Programme d'Investissements d'Avenir'' under the reference ANR-11-IDEX-0004-02. AR acknowledges support from the DGAPA (UNAM) grant IG100218. This research has made use of NASA's Astrophysics Data System.  
\end{acknowledgements}

\bibliographystyle{aa} 
\bibliography{mybibli.bib} 

\begin{appendix}
\section{General equations}
\label{appendixA}
In the analytic part of this work, equations ruling the geometry and the kinematics of a bow shock travelling in a disk wind are given for simplicity in the \textit{"narrow jet"} limit $r_j \to 0$. In this appendix, we give equations valid for an arbitrary jet width that we have used to fit numerical simulations. For the definition of the quantities, we refer to \BT{Figs. \ref{fig:shockframe} and \ref{fig:restframe}}.

\subsection{Shape of the bow-shock and of the cavity}
\BT{Eqs.}~(\ref{mass}) to (\ref{rf}) are valid for an arbitrary width of a jet. Inserting $z_b=z_j-x$ into \BT{Eq.}~(\ref{rx}) we get the the shape of the bow shock ($z_b, r_b$) (see thick cyan line \BT{Fig.~\ref{fig:shockframe}}):
  \begin{equation}
\frac{z_b}{L_0}=  \frac{z_j}{L_0} - \frac{1}{L_0^3} \left( r_b^3  - 3 r_b  r_j^2 + 2 r_j^3 \right).   
\label{zbrb-wide}
\end{equation}
\BT{Eq.}~(\ref{rf}) gives straightforwardly the radius $r_{bf}$ of the edge of the bow shock shell
\begin{equation}
\frac{1}{\gamma L_0^2}\left[r_{b,f}^3-r_j^3+3r_j^2(r_j-r_{b,f})\right]+r_{b,f}-r_j=  \textnormal{v}_0 \textnormal{t}  = z_j \frac{\textnormal{v}_0}{\textnormal{v}_j} \,,
\label{rf-wide}
\end{equation}

Combining \BT{Eqs.}~\ref{zbrb-wide} and \ref{rf-wide} we get the trajectory of the outer edge of the cavity (black dotted line \BT{Fig.~\ref{fig:shockframe}}):
\begin{equation}
\frac{z_f}{L_0} = \frac{\textnormal{v}_w}{\textnormal{v}_j-\textnormal{v}_w} \frac{1}{L_0^3} (r_f^3-3r_f r_j^2 + 2r_j^3) + \frac{\textnormal{v}_j}{\textnormal{v}_0} \frac{1}{L_0} (r_f-r_j).
\label{zf-wide}
\end{equation}

The boundary of the partially refilled cavity (cyan dash-dotted line \BT{Fig.~\ref{fig:shockframe}}) is obtained from \BT{Eqs.}~\ref{rf-wide} and \ref{zf-wide} and is given by:
\begin{equation}
z_c =  \gamma (r_c-r_j) + v_w t .
\label{zc-wide}
\end{equation}
Hence, in the wide jet case, the boundary between the refilled region and the empty cavity has a conical shape.

\subsection{Kinematics}

Integration of \BT{Eqs.}~(\ref{mass}-\ref{xmom}) gives the fully mixed radial and axial velocities:

\begin{equation}
\frac{\textnormal{v}_r}{\textnormal{v}_0}=\left(1+\frac{3(r_b^2-r_j^2)}{\gamma L_0^2}\right)^{-1}\,,
\label{vrv0-wide}
\end{equation}
\begin{equation}
\frac{\textnormal{v}_z}{\textnormal{v}_0}=\frac{\textnormal{v}_w}{\textnormal{v}_0}+\gamma\left(1+\frac{3(r_b^2-r_j^2)}{\gamma L_0^2}\right)^{-1}\,.
\label{vzv0-wide}
\end{equation}

Immediate post-shock velocities obtained by considering the velocity component tangential to the shock surface are:

 \begin{equation}
\textnormal{v}_{r,ps}= (\textnormal{v}_j-\textnormal{v}_w) \frac{3 (r_b^2-r_j^2)/L_0^2}{1+9 \left(r_b^2-r_j^2\right)^2/L_0^4},
\label{vrps-wide}
\end{equation}
\begin{equation}
\textnormal{v}_{z,ps}= \frac{\textnormal{v}_j+ 9 \textnormal{v}_w \left(r_b^2-r_j^2\right)^2/L_0^4}{1+9 \left(r_b^2-r_j^2\right)^2/L_0^4}.
\label{vzps-wide}
\end{equation}

\end{appendix}

\end{document}